\documentclass[a4paper,11pt]{article}
\usepackage[ruled,lined]{algorithm2e}
\usepackage{array}
\newcolumntype{P}[1]{>{\raggedright\arraybackslash}m{#1}}
\usepackage[margin=2.5cm]{geometry}
\usepackage{babel}
\usepackage{hyperref}

\usepackage{amsmath}
\usepackage{amssymb}
\usepackage{multirow}
\usepackage{algorithm2e}
\usepackage{tikz}
\usetikzlibrary{arrows.meta,graphs,shapes.geometric,positioning}
\usepackage{xcolor}

\usepackage[authoryear]{natbib}

\title{A dynamic multi-region MFD model for ride-sourcing with ridesplitting}

\author{Caio Vitor Beojone \\
        \texttt{\href{mailto:caio.beojone@epfl.ch}{caio.beojone@epfl.ch}}
        \and
        Nikolas Geroliminis\thanks{Corresponding author.} \\
        \texttt{\href{mailto:nikolas.geroliminis@epfl.ch}{nikolas.geroliminis@epfl.ch}}
        }

\date{Urban Transport Systems Laboratory (LUTS), \'Ecole Polytechnique F\'ed\'erale de Lausanne (EPFL), Lausanne, CH-1015, Switzerland}

\begin{document}

\maketitle

\begin{abstract}
Dynamic network-level models directly addressing ride-sourcing services can support the development of efficient strategies for both congestion alleviation and promotion of more sustainable mobility.
Recent developments presented models focusing on ride-hailing (solo rides), but no work addressed ridesplitting (shared rides) in dynamic contexts.
Here, we sought to develop a dynamic aggregated traffic network model capable of representing ride-sourcing services and background traffic in a macroscopic multi-region urban network.
We combined the Macroscopic Fundamental Diagram (MFD) with detailed state-space and transition descriptions of background traffic and ride-sourcing vehicles in their activities to formulate mass conservation equations.
Accumulation-based MFD models might experience additional errors due to the variation profile of trip lengths, e.g., when vehicles cruise for passengers.
We integrate the so-called M-model that utilizes the total remaining distance to capture dynamics of regional and inter-regional flows and accumulations for different vehicle (private or ride-sourcing) states.
This aggregated model is capable to reproduce the dynamics of complex systems without using resource-expensive simulations.
We also show that the model can accurately forecast the vehicles' conditions in near-future predictions (e.g., 30 minutes ahead).
Later, a comparison with benchmark models showed lower errors in the proposed model in all states.
Finally, we evaluated the model's robustness to noises in its inputs, and forecast errors remained below 15\% even where inputs were 20\% off the actual values for ride-sourcing vehicles.
The development of such a model prepares the path for developing real-time feedback-based management policies such as priority-based perimeter control or repositioning strategies for idle ride-sourcing vehicles and developing regulations over ride-sourcing in congested areas.

\vspace{11pt}
\end{abstract}

\section{Introduction}

Ride-sourcing became a pronounced service worldwide, providing chauffeured rides to users in mobile applications operated by a Transportation Network Company (TNC) \citep{rayle_etal_2016}.
It is responsible for 15\% of all trips in cities such as San Francisco \citep{sfcta_2017,erhardt_etal_2019}.
Inefficient fleet management may lead to longer vacant travel times and longer passenger waiting times, resembling taxi operations \citep{beojone_geroliminis_2021}.
Such a problem becomes more prominent as the number of ride-sourcing trips is regularly growing.

Ride-sourcing operators might not have a direct interest in congestion (see, for example, \citet{beojone_geroliminis_2021}), but a dynamic model that captures congestion can be valuable for various operational decisions.
Efficient repositioning depends not only on demand knowledge but also on the dynamics of idle vehicles and the time needed to move from one region to the other.
This requires a proper dynamic representation of various states, and we can expand the MFD concept of aggregated modeling in this direction.
Then, if an MFD model could capture critical features of ride-hailing and ridesplitting services, one could investigate various management schemes.

Since the first effort on conceptualizing MFD-based perimeter control \citep{daganzo_2007}, similarly to the approaches of gating, appearing in ramp-metering and signal control, multiple control schemes have been developed.
They include various control methods for heterogeneously congested cities partitioned in homogeneous regions, such as Model Predictive Control (MPC), Proportional-Integral control, optimal control, etc \citep{haddad_shraiber_2014,kouvelas_etal_2017,sirmatel_geroliminis_2018a}.
Models based on the MFD dynamics can forecast near-future conditions of urban systems with lower computational run times than costly simulations, avoiding demanding and detailed route choice and assignment frameworks.
They also require a small number of input data that is more realistic to obtain.
\citet{ramezani_nourinejad_2018} presented the first effort for taxi repositioning control using MFD models, and it had a similar formulation to \citet{geroliminis_2015}, which studied cruising-for-parking.

In most MFD control papers, accumulation-based MFD models are utilized.
These models require the existence of an Outflow-MFD, which considers a unimodal low scatter relation between trip completion and accumulation, assuming a steady-state relationship between production (veh.km travelled per unit time in a region) and outflow that implies a memoryless constant average trip length.
Such an assumption might be problematic for a dynamic ride-sourcing model as the memoryless trip length assumption will not hold.
On the other side, trip-based models (which consider the existence of a Speed-MFD and track the remaining trip length distribution) are computationally demanding and problematic to integrate with control.
A newly developed MFD model, named M-model, provides a decent approximation of trip-based models \citep{murashkin_2021_thesis,lamotte_etal_2018,sirmatel_geroliminis_2021}.
If we can formulate a ride-sourcing model including M-model traffic dynamics, it can be a powerful tool for network-level control, e.g., for relocating vacant vehicles when demand is not balanced, shifting from reactive strategies, where a passenger first leaves unserved so the operator acts to avoid further losses.

Nevertheless, multiple challenges arise when modeling ride-sourcing services with a macroscopic traffic framework.
Firstly, there is a more complex state representation and interactions due to different activities, for examples cruising with no passengers, ride-hailing passengers or ridesplitting passengers, etc.
Namely, the developed framework should be able to provide an understanding of how drivers transition from one activity to another, how these transitions occur in a multi-regional setting, and how they interact with other elements in the traffic system (e.g., passengers and background traffic).
Including a ridesplitting service option not only adds more activities but also adds different dynamics for dealing with passenger-driver matching and how it affects the main movements of drivers.
Especially, serving on a \emph{first-come-first-served} (FCFS) basis requires the understanding that an incoming request can interrupt an ongoing service to assign a new passenger to a shared ride.
Furthermore, passengers have different pick-up and drop-off locations, possibly, all in distinct regions, requiring a driver to deliver the passengers far from each other.
Besides the movement of passengers and drivers in the traffic system, ridesplitting services present significant market thickness in that increases in the demand creates a positive feedback in the service capability of serving multiple passengers.
It all challenges the representation assignments and losses integrated in the same model framework, since macroscopic models do not track individual trips.
Overcoming such challenges is imperative in the described context, where near-future forecasts are essential for developing various managerial frameworks.

Herein, we develop an MFD-based model representing ride-sourcing services and background traffic in a macroscopic multi-region urban network.
The modeled ride-sourcing service offers ride-hailing (single rides) and ridesplitting (shared rides).
Model states describe drivers on their ongoing activities and regions.
We evaluated the proposed model by comparing the errors of the proposed model with benchmarks from the literature utilizing a detailed agent/trip-based simulator developed with real data from the the central business district of Shenzhen, China.
Additionally, a sensitivity analysis investigated the generability of the model to assess the performance of multi-region traffic systems to several service parameters, such as fleet size, willingness to share or waiting time tolerance.
To the best of our knowledge, this is among the first attempts to present and evaluate such a model for ride-sourcing services with the option for ridesplitting (shared rides) in the literature, providing a likely path to the development of strategic repositioning and regulatory problems on ride-sourcing services.

The remainder of the paper has the following structure.
In Section \ref{sec:MFD_basics}, we briefly introduce MFD-based model formulations.
Section \ref{sec:model_description} presents the general modeling framework of the proposed M-model for ride-sourcing with ridesplitting and its particular aspects applied to the modeled service operation.
Section \ref{sec:sensitivity_analysis} evaluates the sensitiveness to service parameters and provide a few managerial insights.
Section \ref{sec:num_results} depicts the numerical results of the proposed model directly compared to the plant and benchmark models and, finally, its robustness to noises in the input data.
Section \ref{sec:conclusions} closes the paper with a discussion of the findings and the final considerations.

\section{MFD modeling basics} \label{sec:MFD_basics}

An MFD provides a well-defined empirical relationship between the number of vehicles in an area (accumulation) and its average speed or the total traveled distance per unit of time (production).
In summary, it can be expressed as $n\mapsto v(n)$ or $n\mapsto P(n)$, where $n$, represents the accumulation, while $v(n)$ and $P(n)$ represent the average space-mean network speed and production, respectively.
Note that one can obtain the $P(n)$ relationship by the product $P(n)=n \cdot v(n)$.
These representations can be called ``speed-MFD'' and ``production-MFD,'' respectively \citep{lamotte_geroliminis_2018}.

On average, all drivers would complete their trips or exit the hypothetical area after traveling a certain distance $L$.
Therefore, assuming that inputs to that area change slower than the relaxation time (time to the travel across the region), the exit function could be written as $O(n)=P(n)/L$ \citep{daganzo_2007}.
Note that an exiting function may represent trips that end (trip-completion) or leave (transfer) the hypothetical area.

Under an input function $\lambda(t)$ and an initial condition $n(0)$, the dynamics of the number of drivers can be described in Equation [\ref{Eq:accum_based_model}], also called the mass-conservation equation \citep{daganzo_2007,mariotte_etal_2017}.
Given the derivation of the exit function, this formulation is usually called the accumulation-based model.

\begin{align}
    \mbox{d}n(t)/\mbox{d}t = \dot{n}(t) = \lambda(t) - O(n(t)) \label{Eq:accum_based_model}
\end{align}

\subsection{Trip-based model}

Consider a driver $m$ with a trip length of $L_m$ entering the system at time $t_m$; then this driver exits the system after travelling $\tau_m$ time units as computed in Equation [\ref{Eq:trip_based_model_base}].

\begin{align}
    L_m = \int_{t_m}^{t_m+\tau_m} v(n(s)) \mbox{ d}s \label{Eq:trip_based_model_base}
\end{align}

Note that the speed $v(n(s))$ results from a speed-MFD.
In other words, the trip-based model considers the traveled distances explicitly, not requiring a particular $L_m=L$ average trip length, resulting in more accurate dynamics for transient situations than the accumulation-based model \citep{paipuri_leclercq_2020}.
However, it becomes difficult to solve analytically, as indicated when introduced in \citet{arnott_2013}.
For this reason, \citet{mariotte_etal_2017} proposed an event-based approach to obtain numerical solutions.
If trip lengths are exponentially distributed, trip-based and accumulation-based models are identical.
Studies mainly propose this model to investigate departure time choice problems at the city scale \citep{arnott_2013,fosgerau_2015,daganzo_lehe_2015,lamotte_geroliminis_2018,jin_2020,vickrey_2020,batista_leclercq_2019,leclercq_paipuri_2019}.

\subsection{Intermediate approach: M-Model}

First introduced in \citet{murashkin_2021_thesis}, the M-model tries to overcome the limitations of the accumulation-based model by summarizing the past events into the total remaining distance $M$ and using this information to update the exit function.
For comparison, trip-based models keep track of individual remaining distances, while accumulation-based models keep no record of it.
Such a model offers valuable intuition and represents an attractive trade-off for control applications.
\citet{sirmatel_tsitsokas_kouvelas_geroliminis_2021} provided a multi-region formulation of the M-model and integrated it successfully in a perimeter control framework.

One can derive the dynamics for computing the total remaining distance based on the dynamics of the accumulation-based model.
Assuming that the average trip length can represent the added remaining distance for each entering driver; then, one can obtain Equation [\ref{Eq:mmodel_remain_dist}].

\begin{align}
    \mbox{d}M/\mbox{d}t = \dot{M}(t) = \dot{n}L = \lambda(t)L - P(n(t)) = \lambda(t)L - n(t)v(n(t)) \label{Eq:mmodel_remain_dist}
\end{align}

To update the exit function, \citet{lamotte_etal_2018} proposed using a correction factor in the form of Equation [\ref{Eq:exit_func_mmodel}].

\begin{align}
    O(t) = \frac{n(t)v(t)}{L}\left(1 + \alpha \left(\frac{M(t)}{n(t)L^*} - 1 \right)\right) \label{Eq:exit_func_mmodel}
\end{align}

\noindent
where $L^*$ is the average remaining distance at steady-state (computed as $L^* = (L^2 + \sigma^2)(2L)^{-1}$, where $\sigma$ is the standard deviation of trip lengths), and $\alpha$ is a constant parameter related to the distribution of trip lengths (see \citet{lamotte_etal_2018} for more details on setting this constant).

Therefore, the M-model uses Equations [\ref{Eq:accum_based_model}] and [\ref{Eq:mmodel_remain_dist}] (replacing the exit function with the result of Equation [\ref{Eq:exit_func_mmodel}]) to keep track of the number of vehicles in the area and their total remaining distance.
For more details about the properties of the M-model, the reader could refer to \citet{murashkin_2021_thesis}.

\section{General model framework} \label{sec:model_description}

In ride-sourcing operations, a TNC uses a platform to centralize trip requests for services, such as ride-hailing and ridesplitting, from incoming passengers and to manage drivers who use their vehicles to profit from offering chauffeured rides \citep{rayle_etal_2016}.
In ride-hailing, single passengers (or a group of related travelers traveling together) request a ride in real-time and the operator tries to assign this trip to nearby drivers who should pick up the passenger and drive to a single destination directly.
Ridesplitting allows multiple unrelated passengers to split rides if their routes overlap.
Therefore, differently from ride-hailing, a detour can occur for at least one of the passengers served by the assigned driver.
Usually, the only additional constraint concerns the added delay/detour to passengers.

Bringing a driver and a passenger together requires a matching process, usually focused on minimizing passengers' waiting times.
If travelers wait too long, they might abandon the trip and use another mode of transport.
For instance, TNCs, such as Uber, try to assign the closest vehicle to a new trip request on a \emph{first-come-first-served} (FCFS) basis \citep{hanna_etal_2016}.
In this paper we make the following assumptions for the matching process of passengers to driver.
Firstly, to maximize the chances of serving ridesplitting requests, the operator can consider interrupting ongoing trips with one passenger so that the vehicle changes its path to deliver both passengers.
That means the operator makes real-time decisions and does not match passengers beforehand nor plans for interruptions.
Secondly, we limit ridesplitting services to at most two simultaneous passengers per vehicle.
Different matching strategies exist in the literature for example, perfect in advance knowledge \citep{santi_etal_2014}, batch matching \citep{alonso_mora_etal_2017} and others \citep{jung_etal_2016,ramezani_nourinejad_2018,berbeglia_etal_2010}.
In this case while different matching processes will require to revisit some aspects of the model, the dynamic framework can still be applied with little extra effort.

The proposed model describes ride-sourcing drivers based on their service assignments, following the provided operation description.
The designed framework also incorporated urban traffic dynamics, tracking private vehicle activities, which formed the majority of background traffic.
Therefore, all drivers in the model fit one of the activities below.

\begin{itemize}
    \item Idle ($I$): a ride-sourcing vehicle with no assignments.
    It is vacant and available for any new passengers.
    \item Ride-hailing ($RH$): a ride-sourcing vehicle with an assigned ride-hailing passenger.
    Or the driver is moving to a pick-up location (Origin) or carrying a ride-hailing passenger towards the destination.
    \item Single ridesplitting ($S1$): a ride-sourcing vehicle with a single assigned ridesplitting passenger.
    Or the driver is moving to an origin or carrying a ridesplitting passenger towards the destination.
    A second ridesplitting assignment can interrupt this service.
    \item Shared ridesplitting ($S2$): a ride-sourcing vehicle that has two ridesplitting passengers assigned.
    Or the driver is moving to one of the pick-up locations or carrying two ridesplitting passengers towards one of their destinations.
    \item Private vehicle ($PV$): a private vehicle (outside the ride-sourcing service) traveling to the destination.
\end{itemize}

\subsection{Macroscopic model dynamics and mass conservation equations}

Besides service-related transitions, drivers will experience different traffic situations while they move on the road network, depending on their current region.
MFD models can describe dynamic state evolution for urban networks partitioned into multiple homogeneously congested regions.
The proposed model uses MFD dynamics to compute the flows of ride-sourcing and private vehicles in a macroscopic urban network.
For illustration, it is composed of a set $\mathcal{R}$ with $R$ heterogeneous regions, i.e., $\mathcal{R}=\{1,2,...,R\}$, each with a well-defined speed-MFD expressing regional speeds as a function of accumulation $v_o(t)=V_o(n_o(t))$.
Therefore, traffic congestion and average speeds are functions of a Speed-MFD, of which the accumulation is the sum of private and ride-sourcing vehicles.
We can scale the function to represent homogeneously congested portions of the area, analogously to \citet{ni_cassidy_2020}.
At the end of the paper, Table \ref{tab:notation} shows a list of notation and definitions.

\begin{figure}[ht]
    \centering
    \begin{tikzpicture}[
statenode/.style={circle, draw=black!70, fill=white!50, thick, minimum size=9mm, inner sep=0pt, font=\footnotesize},
gedgecust/.style={thick, draw=black!70, shorten <=1pt, shorten >=1pt,>={Stealth[round]}},
cube_edge/.style={thick, draw=gray!70, dashed},
pvedgeext/.style={thick, draw=black!70, shorten <=1pt, shorten >=1pt,>={Stealth[round]}, dashed},
]
    \fill[fill=gray!30]
    (-37mm,-8mm) -- (91mm,-8mm) -- (110mm,6mm) -- (110mm,18mm) -- (-18mm,18mm) -- (-37mm,4mm);
    \node[font=\small, text=black!80, rotate=90, above right] at (-37mm, -8mm) {Region $o$};

    \node[statenode]    (priv_OO)   at (-29.0mm,00.0mm)     {$PV_{oo}$};
    \node[statenode]    (idle_OO)   at (29.0mm, 00.0mm)     { $I_{oo}$};
    \node[statenode]    (hail_OO)   at (00.0mm, 00.0mm)     {$RH_{oo}$};
    \node[statenode]    (spt1_OO)   at (58.0mm, 00.0mm)     {$S1_{oo}$};
    \node[statenode]    (spt2_OO)   at (87.0mm, 00.0mm)     {$S2_{oo}$};
    \node[statenode]    (priv_OD)   at (-14.5mm,10.0mm)     {$PV_{od}$};
    \node[statenode]    (idle_OD)   at (43.5mm, 10.0mm)     { $I_{od}$};
    \node[statenode]    (hail_OD)   at (14.5mm, 10.0mm)     {$RH_{od}$};
    \node[statenode]    (spt1_OD)   at (72.5mm, 10.0mm)     {$S1_{od}$};
    \node[statenode]    (spt2_OD)   at (101.5mm,10.0mm)     {$S2_{od}$};
    \node[]             (dummy01)   at (-25.0mm,12.0mm)     {};
    \node[]             (dummy02)   at (-16.0mm,-1.0mm)     {};
    \node[]             (dummy03)   at (-10.0mm,22.0mm)     {};

    \graph[edges=pvedgeext]{ (dummy01)  ->[bend left]  (priv_OO)};
    \graph[edges=pvedgeext]{ (priv_OO)  ->[bend right] (dummy02)};
    \graph[edges=pvedgeext]{ (dummy03)  ->[bend left]  (priv_OD)};
    \graph[edges=gedgecust]{ (priv_OO)  -> (priv_OO)};
    \graph[edges=gedgecust]{ (hail_OO) <-> (idle_OO)};
    \graph[edges=gedgecust]{ (idle_OO) <-> (spt1_OO)};
    \graph[edges=gedgecust]{ (spt1_OO) <-> (spt2_OO)};
    \graph[edges=gedgecust]{ (hail_OD) <-  (idle_OD)};
    \graph[edges=gedgecust]{ (idle_OD)  -> (spt1_OD)};
    \graph[edges=gedgecust]{ (spt1_OD) <-> (spt2_OD)};
    \graph[edges=gedgecust]{ (idle_OO)  -> (hail_OD)};
    \graph[edges=gedgecust]{ (idle_OO)  -> (spt1_OD)};
    \graph[edges=gedgecust]{ (spt1_OO)  -> (spt2_OD)};
    
    \fill[fill=gray!15]
    (-37mm,-38mm) -- (91mm,-38mm) -- (110mm,-24mm) -- (110mm,-12mm) -- (-18mm,-12mm) -- (-37mm,-26mm);
    \node[font=\small, text=black!80, rotate=90, above right] at (-37mm, -38mm)
    {\parbox{26mm}{\raggedright{From Regions $k$ into Region $o$}}};

    \node[statenode]    (priv_KO)   at (-29.0mm,-30.0mm)    {$PV_{ko}$};
    \node[statenode]    (priv_KD)   at (-14.5mm,-20.0mm)    {$PV_{kd}$};
    \node[statenode]    (idle_KO)   at (29.0mm,-30.0mm)     { $I_{ko}$};
    \node[statenode]    (idle_KD)   at (43.5mm,-20.0mm)     { $I_{kd}$};
    \node[statenode]    (hail_KO)   at (00.0mm,-30.0mm)     {$RH_{ko}$};
    \node[statenode]    (hail_KD)   at (14.5mm,-20.0mm)     {$RH_{kd}$};
    \node[statenode]    (spt1_KO)   at (58.0mm,-30.0mm)     {$S1_{ko}$};
    \node[statenode]    (spt1_KD)   at (72.5mm,-20.0mm)     {$S1_{kd}$};
    \node[statenode]    (spt2_KO)   at (87.0mm,-30.0mm)     {$S2_{ko}$};
    \node[statenode]    (spt2_KD)   at (101.5mm,-20.0mm)    {$S2_{kd}$};
    
    \graph[edges=gedgecust]{ (priv_KO)  -> (priv_OO)};
    \graph[edges=gedgecust]{ (priv_KD)  -> (priv_OD)};
    \graph[edges=gedgecust]{ (idle_KO)  -> (idle_OO)};
    \graph[edges=gedgecust]{ (idle_KD)  -> (idle_OD)};
    \graph[edges=gedgecust]{ (hail_KO)  -> (hail_OO)};
    \graph[edges=gedgecust]{ (hail_KD)  -> (hail_OD)};
    \graph[edges=gedgecust]{ (spt1_KO)  -> (spt1_OO)};
    \graph[edges=gedgecust]{ (spt1_KD)  -> (spt1_OD)};
    \graph[edges=gedgecust]{ (spt2_KO)  -> (spt2_OO)};
    \graph[edges=gedgecust]{ (spt2_KD)  -> (spt2_OD)};
    
    \fill[fill=gray!15]
    (-22mm,32mm) -- (106mm,32mm) -- (110mm,36mm) -- (110mm,48mm) -- (-18mm,48mm) -- (-22mm,44mm);
    \node[font=\small, text=black!80, rotate=90, above right] at (-37mm, 22mm)
    {\parbox{26mm}{\raggedright{From Region $o$ into Regions $l$}}};

    \node[statenode]    (priv_LD)   at (-14.5mm,40.0mm)     {$PV_{ld}$};
    \node[statenode]    (idle_LD)   at (43.5mm, 40.0mm)     { $I_{ld}$};
    \node[statenode]    (hail_LD)   at (14.5mm, 40.0mm)     {$RH_{ld}$};
    \node[statenode]    (spt1_LD)   at (72.5mm, 40.0mm)     {$S1_{ld}$};
    \node[statenode]    (spt2_LD)   at (101.5mm,40.0mm)     {$S2_{ld}$};

    \graph[edges=gedgecust]{ (priv_OD) -> (priv_LD)};
    \graph[edges=gedgecust]{ (idle_OD) -> (idle_LD)};
    \graph[edges=gedgecust]{ (hail_OD) -> (hail_LD)};
    \graph[edges=gedgecust]{ (spt1_OD) -> (spt1_LD)};
    \graph[edges=gedgecust]{ (spt2_OD) -> (spt2_LD)};
\end{tikzpicture}
    \caption{General state transition structure.}
    \label{fig:general_state_space}
\end{figure}
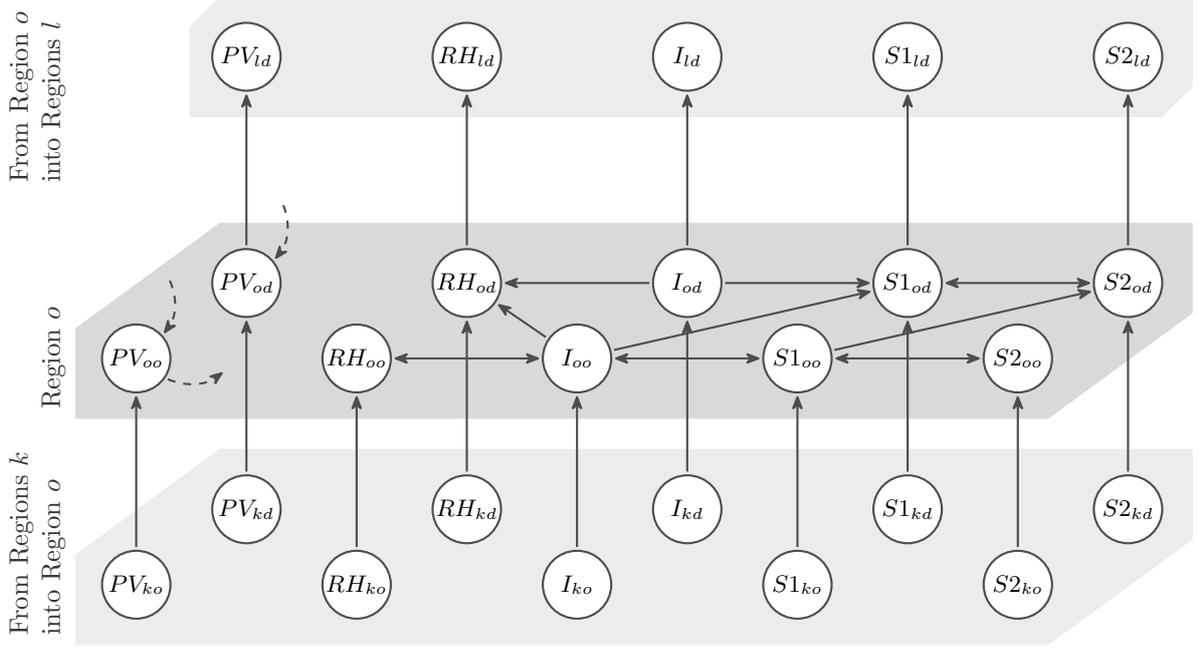

We developed a multi-region M-model which only focuses on vehicular traffic to represent private and ride-sourcing vehicles in different states, which are described based on their activities $K \in \mathbb{A}$ (where $\mathbb{A}= \{I, RH, S1, S2, PV\}$ is the set of activities) and the current and destination regions $od \in \mathcal{R}^2$, summarized into the notation $K_{od}$.
Note that the set of activities $\mathbb{A}$ includes all previously mentioned activities: idle ($I$), ride-hailing ($RH$), single ridesplitting ($S1$), shared ridesplitting ($S2$), and private vehicle ($PV$).
Two sets of conservation equations describe the dynamics of each state (Equations [\ref{Eq:mass_nk}]--[\ref{Eq:mass_MS1}]).
The first one computes the evolution of the number of vehicles, and the second one, of the total remaining distance.
Idle vehicles are the only exception without remaining distance to be estimated because they have no assignments to complete.
Differently than the classical MFD approach, where a vehicle that starts a trip will finalize it with a specific trip length, ridesplitting services contain this additional complexity because of the interruptions.
While a vehicle in an activity $S1$ contributes in the remaining distance of this state with a pre-determined trip length, when a second passenger is assigned the state changes from $S1$ to $S2$ without completing the $S1$ trip, creating an inconsistency in the classical MFD framework (a trip that starts needs to complete its assigned trip length).
That is the reason for $S1_{od}$ being a special case with their particular dynamics accounting for such trip interruptions.
Thus, we only need Equation [\ref{Eq:mass_nk}] to depict idle drivers' dynamics.
In summary, the number of states can be computed as $|\mathcal{K}_{od}|=|\mathcal{R}|+(|\mathbb{A}|-1)\cdot|\mathcal{R}|^2$, where $\mathcal{K}_{od}$ is the set of all states.

\begin{align}
    \dot{n}^K_{od}(t) = & \ \text{Inflow} - \text{Outflow} \hspace{76mm} {K \in \mathbb{A} \backslash S1} \label{Eq:mass_nk} \\
    \dot{M}^K_{od}(t) = & \ \text{Inflow} \cdot \text{Trip length} - n^K_{od}(t)v_o(t) \hspace{50mm} K \in \mathbb{A} \backslash \{I,S1\} \label{Eq:mass_Mk} \\
    \dot{n}^{S1}_{od}(t) = & \ \text{Inflow} - \text{Outflow} - \text{Interruption} \label{Eq:mass_nS1}\\
    \dot{M}^{S1}_{od}(t) = & \ \text{Inflow} \cdot \text{Trip length} - n^{S1}_{od}(t)v_o(t) - \text{Interruption} \cdot \text{Remaining distance} \label{Eq:mass_MS1}
\end{align}

\noindent
where `Inflow', `Outflow' and `Trip length' are defined for each state in Table \ref{tab:support_mass_cons}.
Besides trip lengths, described using the respective $L^K_{od}(t)$, some of the main components of the dynamics $O^K_{od}(t)$ and $O^K_{ohd}(t)$ are trip completion and transfer flow rates, respectively;
where $o$, $h$ and $d$ represent the current, the next and the final region of drivers' path.
On the other hand, $\bar{\lambda}^P_{od}(t)$ (where $P \in \{RH, S1\}$) and $\bar{\lambda}^{S2}_{ohd}(t)$ summarize traveler entering processes assigned to idle drivers or shared ridesplitting rides, respectively;
and the regions $o$, $h$, and $d$ refer to the origin, the intermediate stop of a shared request, and the final destination of an arriving request, respectively.
Private vehicles have their own arrival/enter process for travelers depicted by the value $\bar{\lambda}^{PV}_{od}(t)$.
Equations [\ref{Eq:mass_nk}] and [\ref{Eq:mass_Mk}] did not include vehicles in $S1$ activities due to possible interruptions.
These interfere with the total remaining distance, meaning that part of the production -- total distance traveled per time unit -- does not directly convert into trip completion or transfer flows.
A general state-space framework is illustrated in Figure \ref{fig:general_state_space} focusing on the transitions inside one individual region and its interactions with the neighboring ones, where any region $k$ is preceding region $o$, which is preceding any region $l$ on drivers' path to region $d$ (or region $o$ for those trips ending there).

\newcommand{\tspace}{\rule[-10pt]{0pt}{27pt}}
\begin{table}[ht]
    \centering
    \caption{Summary of dynamic flows in each state (notation described in table 4).}
    \label{tab:support_mass_cons}
    \small
    \begin{tabular}{p{9mm}|p{13mm}|p{14mm}|l|p{36mm}|p{11mm}}
        \hline
        State & Nb. of vehicles & Rem. distance & Inflow & Outflow & Trip length \\
        \hline
        $I_o$ &
            $\displaystyle{n^I_o(t)}$ &
            $\displaystyle{-}$ &
            $\displaystyle{O^{RH}_{oo}(t) + O^{S1}_{oo}(t)}$ &
            $\displaystyle{\sum_{d \in \mathcal{R}}\bar{\lambda}^{RH}_{od}(t) + \bar{\lambda}^{S1}_{od}(t)}$ &
            $\displaystyle{-}$\tspace \\
        $RH_{od}$ &
            $\displaystyle{n^{RH}_{od}(t)}$ &
            $\displaystyle{M^{RH}_{od}(t)}$ &
            $\displaystyle{\bar{\lambda}^{RH}_{od}(t) + \sum_{i \in \mathcal{R}_o}O^{RH}_{iod}(t)}$ &
            $\displaystyle{O^{RH}_{od}(t)}$ &
            $\displaystyle{L^{RH}_{od}(t)}$\tspace \\
        $S1_{od}$ &
            $\displaystyle{n^{S1}_{od}(t)}$ &
            $\displaystyle{M^{S1}_{od}(t)}$ &
            $\displaystyle{\bar{\lambda}^{S1}_{od}(t) + \sum_{i \in \mathcal{R}_o} O^{S1}_{iod}(t) + O^{S2}_{ood}(t)}$ &
            $\displaystyle{O^{S1}_{od}(t)}$ &
            $\displaystyle{L^{S1}_{od}(t)}$\tspace \\
        $S2_{od}$ &
            $\displaystyle{n^{S2}_{od}(t)}$ &
            $\displaystyle{M^{S2}_{od}(t)}$ &
            $\displaystyle{\text{Interruptions} + \sum_{i \in \mathcal{R}_o} O^{S2}_{iod}(t)}$ &
            $\displaystyle{O^{S2}_{od}(t)}$ &
            $\displaystyle{L^{S2}_{od}(t)}$\tspace \\
        $PV_{od}$ &
            $\displaystyle{n^{PV}_{od}(t)}$ &
            $\displaystyle{M^{PV}_{od}(t)}$ &
            $\displaystyle{\bar{\lambda}^{PV}_{od}(t) + \sum_{i \in \mathcal{R}_o} O^{PV}_{iod}(t)}$ &
            $\displaystyle{O^{PV}_{od}(t)}$ &
            $\displaystyle{L^{PV}_{od}(t)}$\tspace \\
        \hline
    \end{tabular}
\end{table}

Note that there is no endogenous trip completion for idle $I_o$ drivers since they do not have any assignments to complete; instead, they cruise for passengers, and they exit their current state through passenger entering processes.
While one could compute \emph{a posteriori} what is the average distance traveled for vehicles in state $I_o$, this is not defined in the classical way as in MFD models because it is state-specific and varies over time.

In Table \ref{tab:support_mass_cons}, inflows and outflows of state $RH_{od}$ illustrate that drivers do not deliver their passengers before entering the destination region.
Every new ride-hailing assignment adds its average trip length to the remaining distance.
The model assumes, without loss of generality, a single average trip length for all input flows.\footnote{%
To relax this assumption, one must use individual trip lengths for each of the entering possibilities in the state.
Then, the computation of $O^K_{od}(t)$ can later use a weighted average (based on the input rates) of these trip lengths.}
Later, in Section \ref{sec:trip_lengths} the calculation of every trip length is detailed.

Most inflows and outflows for ridesplitting are naturally compatible with those for ride-hailing.
However, the traveler entering process has a double role.
While a portion serves as inflow for state $S1_{od}$, the remaining interrupts a current service, as an additional passenger is entering the vehicle.
Traditionally, in MFD-based models, vehicles must always complete the started trips.
Interruptions violate such an assumption.
Therefore, Equations [\ref{Eq:mass_nS1}] and [\ref{Eq:mass_MS1}] relax this assumption for state $S1_{od}$.

In Equation [\ref{Eq:interruption}], we separate the interruptions depending on the trip scheme resulting from the matching process, namely \emph{last-in-first-out} (LIFO) and \emph{first-in-first-out} (FIFO), where the last and first refer to the passengers entering and exiting the vehicle.
Note that $\bar{\lambda}^{S2}_{ohd}(t)$ and $\bar{\lambda}^{S2}_{odh}(t)$ represent the rate of ridesplitting requests assigned into a LIFO and a FIFO shared ridesplitting trip-schemes, respectively.
The first part, where we have $\sum_{h \in \mathcal{R}} \bar{\lambda}^{S2}_{ohd}(t)$, refers to cases where the destination of the new traveler lies in one of the possible regions on the path of the initial trip, being, thus, delivered earlier than the initial passenger (LIFO trip scheme).
The second part, where we have $\sum_{h \in \mathcal{R}} \bar{\lambda}^{S2}_{odh}(t)$, refers to cases where the destination of the new traveler is farther than the initial one, being, thus, delivered after the initial passenger (FIFO trip scheme).
In LIFO trip schemes, drivers in state $S1_{od}$ will enter a state $S2_{od}$ for an incoming $oh$ request.
In FIFO trip schemes, drivers in state $S1_{od}$ will enter a state $S2_{oh}$ for an incoming $oh$ request.
Note that there must be a similarity between the $od$ and $oh$ so that $\bar{\lambda}^{S2}_{ohd}(t)>0$ or $\bar{\lambda}^{S2}_{odh}(t)>0$.
We describe in detail how to estimate these values later in Section \ref{sec:particular_model}, for now it stands for the general framework of the proposed model.

\begin{align}
    \text{Interruptions} = & \underbrace{\sum_{h \in \mathcal{R}} \bar{\lambda}^{S2,\text{LIFO}}_{ohd}(t)}_{\text{LIFO trip scheme}} + \underbrace{\sum_{h \in \mathcal{R}\backslash\{d\}} \bar{\lambda}^{S2,\text{FIFO}}_{odh}(t)}_{\text{FIFO trip scheme}} \label{Eq:interruption}
\end{align}

\noindent
We only accounted for $\bar{\lambda}^{S2}_{odd}(t)$ in the LIFO trip scheme to avoid double counting.

The last element of Equation [\ref{Eq:mass_MS1}] is the `Remaining distance' $L^*_{S1_{od}}(t)$ at the moment of the second assignment (interruption).
It illustrates the effect of not completing a trip as initially planned, while the `Interruption' illustrates the process.

Recall that the entering flow of passengers in activity $S2$ is equivalent to the interruptions in activity $S1$.
Drivers in $S2_{od}$ carry two passengers with possibly different destination regions but with a similar regional path.
Therefore, these drivers might drop a passenger before reaching $d$.
This results from the possible trip schemes in shared rides.
The trip length $L^{S2}_{od}(t)$ has no relation to the remaining distance $L^*_{S1_{od}}(t)$.

Finally, throughout the paper, we refer to trip length as the space a driver travels in a region (focus of the traffic model), not the whole distance that a driver covers from the assignment until the drop-off area.
Although related, they represent different aspects of the ride-sourcing operation in a traffic model.
While the first one relates to drivers' network movements, the second one relates to individual service requests.
We should mention that even though some regional trip lengths are the same for different services/activities, the total trip length from the time of a vehicle assigned to the first passenger, until completing the trip varies across time and sequence of activities.

\subsection{Trip completion rates, transfer flows} \label{sec:trip_compl}


Once assigned, the driver enters a busy state and starts one of the outflow processes of trip completion or transfer.
In terms of notation, the difference between them is the state description, where state $K_{od}$ completes a trip if $o=d$ or transfers if $o \neq d$.
Vehicles outside their region of destination must first transfer along the trip path and then complete the trip.
The proposed M-model computes outflows using Equation [\ref{Eq:outflow}].
Recall that $I_o$ drivers' outflow is the passenger entering process.

\begin{align}
    O^K_{od}(t) = \frac{n^K_{od}(t)v_o(t)}{L^K_{od}(t)} \left(1 + \alpha \left(\frac{M^K_{od}(t)}{n^K_{od}(t)L^*_{K_{od}}} - 1\right) \right), && K \in \mathbb{A} \backslash \{I,S1\}, \mbox{ and } \{o,d\} \in \mathcal{R}^2 \label{Eq:outflow}
\end{align}

\noindent
where $O^K_{od}(t)$ is the instantaneous trip completion/transfer rate for vehicles in state $K_{od}$;
$\alpha$ is a model parameter expressing the sensitivity of outflow to variations in the remaining distance $M^K_{od}(t)$.
$L^K_{od}$ is the average trip length.
$L^*_{K_{od}}$ is the steady-state average remaining distance in state $K_{od}$ until exiting the current region.
It can be computed as a function of the average trip length $L^K_{od}(t)$ and its standard deviation $\sigma^K_{od}(t)$, i.e., $L^*_{K_{od}}(t)=\left(L^K_{od}(t)^2+\sigma^K_{od}(t)^2\right)\left(2L^K_{od}(t)\right)^{-1}$.

Recall that the interruption of an ongoing $S1_{od}$ delivery for a new shared ride leaves a $L^*_{S1_{od}}(t)$ distance to the destination or next region uncovered, meaning that part of the production of these vehicles does not convert into trip completion or transfer rates.
We use mass conservation Equations [\ref{Eq:mass_nS1}] and [\ref{Eq:mass_MS1}] at steady-state to estimate an approximation of $O^{S1}_{od}(t)$.
The idea in Equations [\ref{Eq:proof_OS1_n0}]--[\ref{Eq:proof_OS1_m0}] is to isolate the inflow to combine both equations and obtain the estimate in Equation [\ref{Eq:proof_OS1_end}].

\begin{align}
    \dot{n}^{S1}_{od}(t) & = \text{Inflow} - O^{S1}_{od}(t) - \text{Interruption} = 0 \Rightarrow \nonumber\\
    \Rightarrow \text{Inflow} & = O^{S1}_{od}(t) + \text{Interruption} \label{Eq:proof_OS1_n0}\\
    \dot{M}^{S1}_{od}(t) & = \text{Inflow} \cdot L^{S1}_{od}(t) - n^{S1}_{od}(t)v_o(t) - \text{Interruption} \cdot L^*_{S1_{od}}(t) = 0 \Rightarrow \nonumber \\
    \Rightarrow \text{Inflow} & = \frac{n^{S1}_{od}(t)v_o(t) + \text{Interruption} \cdot L^*_{S1_{od}}(t)}{L^{S1}_{od}(t)} \label{Eq:proof_OS1_m0} \\
    O^{S1}_{od}(t) & = \frac{n^{S1}_{od}(t)v_o(t) + \text{Interruption} \cdot L^*_{S1_{od}}(t)}{L^{S1}_{od}(t)} - \text{Interruption} \nonumber \\
     & = \frac{n^{S1}_{od}(t)v_o(t)}{L^{S1}_{od}(t)} - \left(1 - \frac{L^*_{S1_{od}}(t)}{L^{S1}_{od}(t)}\right) \cdot \text{Interruption} \nonumber \\
     & = \hat{O}^{S1}_{od}(t) - \left(1 - \frac{L^*_{S1_{od}}(t)}{L^{S1}_{od}(t)}\right) \cdot \text{Interruption} \label{Eq:proof_OS1_end}
\end{align}

\noindent
where $L^{S1}_{od}(t)$ is the trip length of newly assigned single ridesplitting trips;
$\hat{O}^{S1}_{od}(t)$ is an estimator of the trip completion rate (or transfer flow) without interruptions;
and the second term of the result is the amount of $\hat{O}^{S1}_{od}(t)$ to be discounted due to interruptions.
Finally, one can estimate $\hat{O}^{S1}_{od}(t)$ using the M-model approximation from Equation [\ref{Eq:outflow}].

Drivers may drive through different routes between their current and destination regions.
The outflow $O^K_{ohd}(t)$ is the transfer rate from a current region $o$ through the immediate next one $h$, as illustrated in Equation [\ref{Eq:outflow_neighbor}].
Therefore, $\theta_{ohd} \in [0,1]$ distributes transfer flows over its neighboring regions such that the equality $\sum_{h \in \mathcal{R}_o} \theta_{ohd} = 1$ holds.
The internal trip completion rate is computed directly from Equation [\ref{Eq:outflow}], where $o=d$.

\begin{align}
    O^K_{ohd}(t) = & \ \theta_{ohd} \! \cdot \! O^K_{od}(t) & K \in \mathbb{A} \backslash \{S2\}\label{Eq:outflow_neighbor}
\end{align}

Drivers in $S2_{od}$ are an exception to the previous because they may deliver one of their passengers in a region before the last destination.
The process for dropping one of the passengers in the current region precedes transfer.
Hence, part of the drivers will transfer (Equation [\ref{Eq:outflow_S2_transfer}]), while others will return to state $S1_{od}$ before transferring (Equation [\ref{Eq:outflow_S2_drop}]).

\begin{align}
    O^{S2}_{ohd}(t) = & \ \theta_{ohd} \! \cdot \! (1 - \vartheta_{ood}(t)) \! \cdot \! O^{S2}_{od}(t) & h \in \mathcal{R}_o \label{Eq:outflow_S2_transfer} \\
    O^{S2}_{ood}(t) = & \ \vartheta_{ood}(t) \! \cdot \! O^{S2}_{od}(t) \label{Eq:outflow_S2_drop}
\end{align}

\noindent
where $\vartheta_{ood}(t)$ becomes the fraction of shared trips passing through $o$ that will deliver a passenger before continuing to $d$.

Regional trip information is essential for MFD-based models, and, in the case of ride-sourcing, it results from the passenger-driver matching process.
Ride-sourcing drivers pick up and then deliver their assigned passengers and Equation [\ref{Eq:trip_length}] breaks trip lengths into both activities.
Recalling that, trip lengths in the proposed model are the distances traveled in a region, not the ones for whole assignments.
In this paper we assume that the values for $\theta_{ohd}$ are exogenously defined, while there are approaches in the MFD literature to integrate through an aggregated assignment process, which is beyond the scope of this work (see for example \citet{yildirimoglu_geroliminis_2014}).

\begin{align}
    L^K_{od}(t)=L^\text{pick}_{K_{od}}(t)+L^\text{drop}_{K_{od}}(t) \label{Eq:trip_length}
\end{align}

\subsection{Drivers movements and passenger-driver matching process} \label{sec:particular_model}


Matching passengers with available vehicles is at the core of ride-sourcing operations.
It defines if an arriving passenger enters the system or leaves unserved.
Replicating it in a dynamic model requires translating such a microscopic activity into a macroscopic scheme.
We intended to identify a parsimonious way to integrate matching in an aggregated model.

Most efforts of replicating this process into a dynamic model use Cobb-Douglas matching friction function.
Examples of such include \citet{ramezani_nourinejad_2018}, \citet{xu_etal_2020b}, and \citet{nourinejad_ramezani_2020}.
However, it assumes that passengers wait for an amount of time and form a batch to assign them to drivers, such that it balances demand and the supply of drivers.
None of the previous studies handled large-scale ride-sourcing with a ridesplitting option with an FCFS assignment.

As a consequence of the FCFS assignment, we can formulate a loss probability function in response to endogenous variables such as
the available fleet size $n^{av}_{od}(t)$,
the regional average speed $v_o(t)$,
the waiting time tolerance $\omega$, and
the ratio of Idle-Busy drivers $\rho^s_{od}$.\footnote{%
We must highlight that this approximation may be revised in the case of the matching process batch passengers before the assignment and cases where the supply of drivers is too small.
For instance, if one desires to batch passengers, instead of loss probabilities, one can create a state of waiting passengers in each region, and use a Cobb-Douglas function or an input-output diagram to compute the number of matches after batching some passengers and drivers (as already seen in the literature before).
Furthermore, to integrate multiple degrees of patience, one can group passenger arrival processes according to their waiting time tolerances.\label{foot:batch}}
It constitutes an operational result for a given demand profile and response to service quality requirements.
The function should yield a few properties that will later ensure non-negativity to drivers' numbers and steer the consequences of the matching process, including the thickness to demand.
Firstly, $pl^s_{od} \in (0,1]$ for $n^{av}_{od},v_o,\omega,\rho^s \geq 0$.
Secondly, if any of the parameters approaches $0$, then $pl^s_{od}$ approaches $1$.
Thirdly, all partial derivatives are negative, i.e., all parameters decrease the chances of losing incoming requests.
With these assumptions, a Cobb-Douglas type function fits the negative log probability to ensure the previously mentioned desired properties.
Equation [\ref{Eq:loss_prob_form}] depicts the formulation.

\begin{align}
    pl^s_{od}(n^{av}_{od},v_o,\omega,\rho^s_{od}) = & \ \exp \left(-\gamma_0 \cdot (n^{av}_{od})^{\gamma_1} \cdot (v_o)^{\gamma_2} \cdot \omega^{\gamma_3} \cdot (\rho^s_{od})^{\gamma_4} \right) \label{Eq:loss_prob_form}
\end{align}

All parameters $\gamma_q, q\in\{0,1,2,3,4\}$ must be positive to ensure negative partial derivatives.
One must acknowledge that $\gamma_0$ represents the coverage of a vehicle, while $\gamma_1$, $\gamma_2$, $\gamma_3$, $\gamma_4$ indicate the coverage sensitivity to each endogenous variable.
Furthermore, the these parameters are not dependent on the service option.
However, endogenous variables can vary between service options depending on particular aspects of the modeled system.
We have to note that these curves might not be universal and depend on the matching policy.

Note that the computation of $pl^s_{od}$ considered all available drivers $n^{av}_{od}(t)$ because one cannot identify which drivers would be capable of serving the arriving request before the assignment.
Therefore, if we account only for drivers that would comply with all the constraints, we would be changing the sample space, in an example of the ``Monty Hall'' problem.

A Monte Carlo simulation followed by a linear regression model evaluated the parameters $\gamma_q, q \in \{0,1,2,3,4\}$ of Equation [\ref{Eq:loss_prob_form}] to compute the loss of incoming requests after checking their feasibility constraints (waiting time and/or detour).
Appendix \ref{sec:monte_carlo} details the construction and algorithm of the Monte Carlo simulation, while Appendix \ref{sec:linear_model} summarizes how Equation [\ref{Eq:loss_prob_form}] was linearized to construct a linear regression model with the results of the Monte Carlo simulation.
For instance, the linear regression of a single region experiment with the entire road network obtained a $R^2=0.96$.

Since ride-sourcing cannot serve all arriving customers, exogenous arrival rates enter the ride-sourcing restrained by the respective loss probabilities $pl^s_{od}(t)$ (simpler notation of Equation [\ref{Eq:loss_prob_form}]) in Equation [\ref{Eq:arr_RS}].
We assume that lost customers use private vehicles, so we penalize the congestion and maintain the total number of trips.
Equation [\ref{Eq:arr_PV}] adds these lost requests to the private vehicle demand.

\begin{align}
    \bar{\lambda}^s_{od}(t) = & \left(1-pl^s_{od}(t)\right)\lambda^s_{od}(t), && \mbox{where, } s \in \{H,S\}  \label{Eq:arr_RS} \\
    \bar{\lambda}^{PV}_{od}(t) = & \lambda^{PV}_{od}(t) + \sum_{s \in \{H,S\}} pl^s_{od}(t)\lambda^s_{od}(t) \label{Eq:arr_PV}
\end{align}

\noindent
where $\lambda^s_{od}(t)$ and $\lambda^{PV}_{od}(t)$ are the arrival rate of travelers for one of the ride-sourcing services (ride-hailing $H$ or ridesplitting $S$) and private vehicles, respectively.
Then, $\bar{\lambda}^s_{od}(t)$ and $\bar{\lambda}^{PV}_{od}(t)$ are the entrance rate of these travelers (we distinguish between arrival and entrance).

The proposed model assumes a similar geographical distribution among all available drivers for a service.
Therefore, Equation [\ref{Eq:gamma}] can endogenously compute the proportion of entering passengers assigned to idle drivers, $\rho^s_{od}(t)$, based on the instantaneous number of drivers;
while the probability of assigning it to a busy vehicle is $1-\rho^s_{od}(t)$.\footnote{%
One can extend the ratios $\rho^s_{od}(t)$ to larger passenger capacities if the assumptions remain the same for drivers' geographical distribution and dispatching policy.}$^{,}$\footnote{%
Cases with distinct dispatching policies may require direct prioritization of certain vehicles (such as priority to vehicles at state $S1$), requiring adaptations to the computation of the  $\rho^s_{od}(t)$ and $pl^s_{od}(t)$.}
The previous assumes that available vehicles and arriving passengers must be in the same region.
Such an assumption is reasonable if the number of assigned drivers across the region limits is negligible.
The number of available drivers for a shared ridesplitting ride differs, depending on the evaluated trip-scheme.
Therefore, Equations [\ref{Eq:nAV_LIFO}] and [\ref{Eq:nAV_FIFO}] counts the number of available drivers for LIFO and FIFO trip-schemes, respectively.

\begin{align}
    \rho^s_{od} (t) = \frac{n^I_o(t)}{n^{av}_{od}(t)} = & \ 
    \begin{cases}
        \displaystyle{\frac{n^I_o(t)}{n^I_o(t)}=1}, & \text{if } s = H \\[10pt]
        \displaystyle{\frac{n^I_o(t)}{n^I_o(t) + n^{av,\text{LIFO}}_{od}(t) + n^{av,\text{FIFO}}_{od}(t)}}, & \text{if } s = S
    \end{cases}
    \label{Eq:gamma} \\[10pt]
    n^{av,\text{LIFO}}_{od}(t) = & \sum_{h \in \mathcal{R}} \beta_{oh}^d n^{S1}_{oh}(t) \label{Eq:nAV_LIFO} \\
    n^{av,\text{FIFO}}_{od}(t) = & \sum_{h \in \mathcal{R}\backslash\{d\}} \beta_{od}^h n^{S1}_{oh}(t) \label{Eq:nAV_FIFO}
\end{align}

\noindent
where $\beta_{od}^h$ ($\beta_{oh}^d$) represents the ratio of $od$ ($oh$) trips that will pass through region $h$ ($d$).

The computation of the loss probability $pl^s_{od}(t)$ considers that unacceptable detours will restrain some of the demand from entering the service (see the Appendix \ref{sec:monte_carlo} for details).
Therefore, one should not limit available drivers $n^{av}_{od}(t)$ to those complying with all constraints (wait and detour), under the penalty of accounting twice for the same effects.

We can further divide the passenger entrance process according to the activity assigned to the driver.
In the case of ride-hailing, Equation [\ref{Eq:enter_RH}] confirms that entering passengers causes idle drivers to enter state $RH_{od}$.
For ridesplitting, Equation [\ref{Eq:enter_S1}] states the process of having drivers assigned to single requests.
Equations [\ref{Eq:enter_S2_LIFO}] and [\ref{Eq:enter_S2_FIFO}] illustrate the assignment of drivers to shared ridesplitting requests in LIFO and FIFO trip schemes, respectively.

\begin{align}
    \bar{\lambda}^{RH}_{od}(t) = & \ \rho^H_{od}(t) \bar{\lambda}^H_{od}(t) = \bar{\lambda}^H_{od}(t) \label{Eq:enter_RH} \\
    \bar{\lambda}^{S1}_{od}(t) = & \ \rho^S_{od}(t)\bar{\lambda}^S_{od} \label{Eq:enter_S1} \\
    \bar{\lambda}^{S2,\text{LIFO}}_{ohd}(t) = & \ \left(1 - \rho^S_{oh}(t)\right) \! \cdot \! \frac{\beta_{od}^h n^{S1}_{od}(t)}{n^{av,\text{LIFO}}_{oh}(t) + n^{av,\text{FIFO}}_{oh}(t)} \bar{\lambda}^S_{oh}(t) \label{Eq:enter_S2_LIFO} \\
    \bar{\lambda}^{S2,\text{FIFO}}_{odh}(t) = & \ \left(1 - \rho^S_{oh}(t)\right) \! \cdot \! \frac{\beta_{oh}^d n^{S1}_{od}(t)}{n^{av,\text{LIFO}}_{oh}(t) + n^{av,\text{FIFO}}_{oh}(t)} \bar{\lambda}^S_{oh}(t) && h \neq d \label{Eq:enter_S2_FIFO}
\end{align}

In the assumed matching process, one of the passengers will be delivered first in shared ridesplitting.
Equation [\ref{Eq:theta_ood}] uses current demand information to identify the fraction of shared trips delivering a passenger in their current region, $\vartheta_{ood}(t)$.\footnote{%
Tracking every delivery stop in the state notation would relax this memoryless assumption by deteriorating model's scalability.}

\begin{align}
    \vartheta_{ood}(t) = & \ \frac{\sum_{h \in \mathcal{R}}\bar{\lambda}^{S2}_{hod}(t)}{\sum_{h \in \mathcal{R}} \sum_{l \in \mathcal{R}} \beta_{hl}^o \beta_{od}^l \bar{\lambda}^{S2}_{hld}(t)} \label{Eq:theta_ood}
\end{align}

\noindent
where, $\sum_{h \in \mathcal{R}} \sum_{l \in \mathcal{R}} \beta_{hl}^o \beta_{od}^l \bar{\lambda}^{S2}_{hld}(t)$ indicates all demand for shared ridesplitting trips heading to region $d$ that will pass through $o$ (either for delivering a passenger, or just as a passage towards $d$) before delivering one of the passengers.

\subsection{Trip length estimates} \label{sec:trip_lengths}

In assuming a FCFS matching process to the nearest available driver, pick-up trip lengths become analogous to the average minimal distance to the center of a circle.
\citet{daganzo_2010_book} and \citet{daganzo_ouyang_2019_book} derived the approximation in Equation [\ref{Eq:dist_pick}], where the product between waiting time tolerance $\omega$ and instantaneous speed $v_o(t)$ determines the matching radius.
The literature presents other similar results \citep{zhang_etal_2019b,zhang_nie_2019}.
From the loss probability, we approximate the number of matchable drivers (number of drivers that are available and comply with all matching constraints) as $(1-pl^s_{od}(t))n^{av}_{od}(t)$.

\begin{align}
    L^\text{pick}_{K_{od}}(t) \approx 0.63\sqrt{\frac{\omega \cdot v_o(t)}{(1-pl^s_{od}(t)) \cdot n^{av}_{od}(t)}} \label{Eq:dist_pick}
\end{align}

We averaged time invariant trip lengths for delivery, $L^\text{drop}_{K_{od}}$, based on historical data.
The previous refers only to intra-regional trip lengths in the historical data, not the length of multi-region trips, which would vary according to the demand (Origin-Destination combination) and vehicle routing choices.

The model also uses the remaining distance $L^*_{K_{od}}$ in outflows (trip completion and transfer flows) and interruptions.
As mentioned earlier, in steady-state $L^*_{K_{od}}(t) = \left(L^K_{od}(t)^2+\sigma^K_{od}(t)^2\right) \! \cdot \! \left(2L^K_{od}(t)\right)^{-1}$.
Demand changes may take the traffic system out of a steady-state condition, changing the actual value of $L^*_{K_{od}}$.

The coefficient of variation ($\sigma/L$) remained almost constant in historical data (computed from several simulations, with the data aggregated in time intervals of 3 minutes for each activity), ranging between 0.54 and 0.64 (depending on the experimental settings in the simulation, OD-pair and number of regions), which is far from a value of 1, justifying the choice of an M-model.
Furthermore, given the nearly constant coefficients of variation, there is no need to compute $\sigma^K_{od}(t)$ separately, simplifying the computation of $L^*_{K_{od}}$.

\section{Model's sensitivity analysis in a multi-region setting} \label{sec:sensitivity_analysis}

\begin{figure}[ht]
    \centering
    \includegraphics{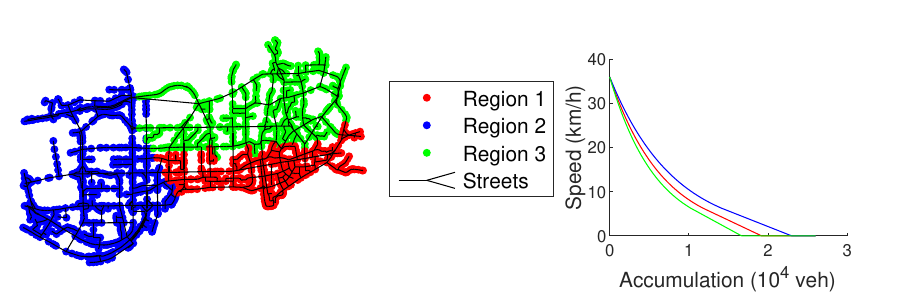}
    \caption{Shenzhen central business district separated in three regions and their respective Speed-MFDs used in the sensitivity analysis.}
    \label{fig:3reg_speedMFD}
\end{figure}

A multi-region setting can provide insights for those interested in developing regulatory schemes envisioning better traffic conditions and general welfare.
Nevertheless, the platform operator can also derive rapid forecasts and evaluate possible decisions and near-future consequences in the operation of the service.
Therefore, in this section, we assume a three-region model described with the equations of Section \ref{sec:model_description} and we investigate the dynamic evolution of states for different fleet size, willingness to share and waiting time tolerance with the parameters depicted in Table \ref{tab:3reg_data}.
We computed the data based on the demand data from Shenzhen \citep{bellocchi_geroliminis_2020}.
Other data include a constant coefficient of variation $\sigma/L=0.650$ and the Speed-MFD, as depicted in Figure \ref{fig:3reg_speedMFD}.

\begin{table}[ht]
    \centering
    \caption{Three-region model parameters.}
    \begin{tabular}{ccccccccccc}
        \hline
        \multirow{2}{*}{OD-pair} & \multirow{2}{15mm}{\centering{Demand ratio}} & \multicolumn{3}{c}{Transfer ratios ($\theta_{okd}$)} & \multicolumn{3}{c}{Passage ratio ($\beta^k_{od}$)} & \multicolumn{3}{c}{$L^\text{drop}_{K_{od}}$ (km)} \\
         & & $k=1$ & $k=2$ & $k=3$ & $k=1$ & $k=2$ & $k=3$ & $RH/PV$ & $S1$ & $S2$ \\
        \hline
        1-1 & 0.136 & n/d & n/d & n/d   & 1 & 0 & 0     & 2.119 & 2.084 & 2.144 \\
        1-2 & 0.070 & 0 & 0.916 & 0.084 & 1 & 1 & 0.084 & 2.805 & 2.860 & 2.562 \\
        1-3 & 0.099 & 0 & 0.005 & 0.995 & 1 & 0.004 & 1 & 1.861 & 2.059 & 1.923 \\
        2-1 & 0.061 & 0.904 & 0 & 0.096 & 1 & 1 & 0.095 & 3.257 & 3.307 & 2.758 \\
        2-2 & 0.308 & n/d & n/d & n/d   & 0 & 1 & 0     & 3.128 & 3.148 & 3.074 \\
        2-3 & 0.055 & 0.535 & 0 & 0.465 & 0.534 & 1 & 1 & 3.304 & 3.079 & 3.015 \\
        3-1 & 0.091 & 0.995 & 0.005 & 0 & 1 & 0.005 & 1 & 2.095 & 2.168 & 1.855 \\
        3-2 & 0.054 & 0.449 & 0.551 & 0 & 0.449 & 1 & 1 & 2.820 & 2.807 & 2.579 \\
        3-3 & 0.126 & n/d & n/d & n/d   & 0 & 0 & 1     & 2.292 & 2.297 & 2.274 \\
        \hline
        \multicolumn{11}{l}{(n/d: non-defined)}
    \end{tabular}
    \label{tab:3reg_data}
\end{table}

We set the plant to reproduce dynamic traffic entering the hyper-congested regime and then returning to an uncongested state without reaching gridlock.
Such a setting is supposed to generate a challenging scenario for the model evaluation since hyper-congested situations create conditions far from steady-state and hysteresis during the loading and unloading of the network.
A Poisson process describes the arrivals in piece-wise constant rates during 3 hours of simulation such that there is a peak hour preceded and followed by low-demand hours.
It generates a total of 40,000 trips per hour during the low-demand hours and 70,000 trips per hour during the peak hour, from which 85\% are background traffic and 15\% are ride-sourcing requests.

\begin{figure}[ht]
    \centering
    \includegraphics{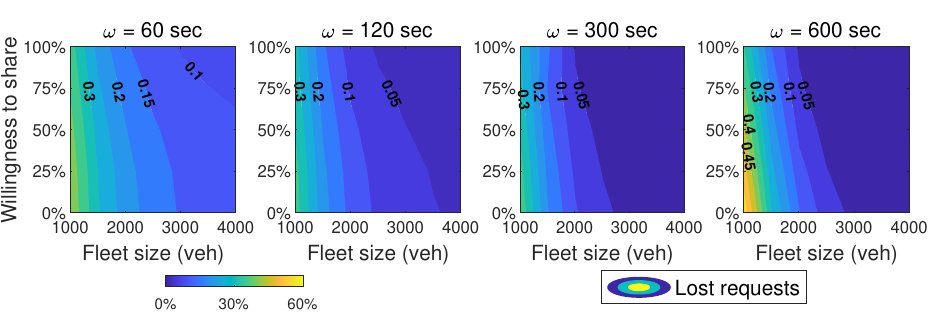}
    \caption{Summary of abandonment rates as a function of fleet size, willingness to share and waiting time tolerance ($\omega$).}
    \label{fig:multi_reg_abandonment}
\end{figure}

Firstly, passengers may have different tolerances and service preferences, while the platform can manage its service to influence the number of active drivers and passengers' service choices.
Therefore, in Figure \ref{fig:multi_reg_abandonment}, we quantify the effects of passengers' willingness to share (i.e., the fraction of ride requests for ridesplitting), their waiting time tolerance $\omega$, and the fleet size of active ride-sourcing drivers.
As expected, increased fleet sizes and willingness to share decreased the fraction of lost requests.
For instance, with a waiting time tolerance of 60 seconds, a fleet size of about 2100 drivers reaches the same 15\% abandonment ratio as a fleet of 2800 drivers when the willingness to share increases from 25\% to 100\%.
However, waiting time tolerance creates a different behavior.
On the one hand, more patient passengers enlarge the coverage area for pick-up, increasing the chances for passenger-driver matching.
On the other hand, it allows assigning drivers farther from their passengers, which keeps them busy for prolonged periods, decreasing their availability for incoming requests, in one consequence of the wild-goose chase effect \citep{castillo_etal_2018}.

\begin{figure}[ht]
    \centering
    \includegraphics{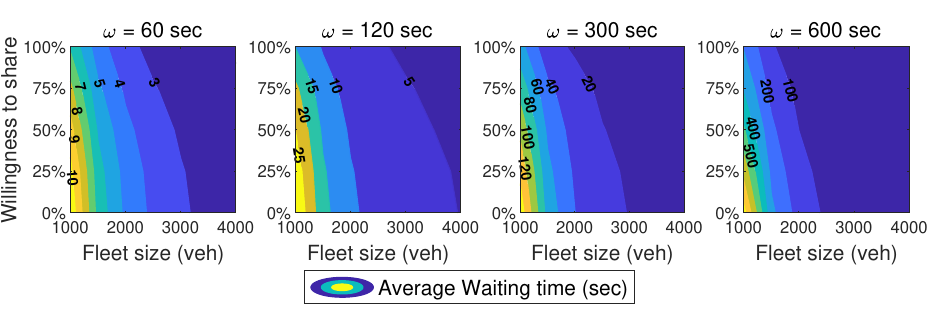}
    \caption{Average waiting time (including a penalty for abandonment) as a function of fleet size, willingness to share and waiting time tolerance ($\omega$).}
    \label{fig:multi_reg_waiting}
\end{figure}

As a direct consequence of the same settings, they affect the average waiting time, a key performance indicator to attract and maintain customers in this service.
In Figure \ref{fig:multi_reg_waiting}, passengers' waiting time tolerance causes the most significant changes in waiting time, changing average values in orders of magnitude from less than 10 seconds to near 10 minutes for tolerances of 1 and 10 minutes, respectively.
Note that these numbers are also affected by an abandonment penalty, such that the average waiting time increases by the abandonment rate as $\text{Penalized waiting time} = \text{Waiting time} \cdot (1 + \text{Abandonment})$ \citep{beojone_geroliminis_2021}.
Enlarging the available fleet had higher impacts than passengers' willingness to share.
Pairing both fleet size and willingness to share can achieve more efficient outcomes, such as keeping the same 20 seconds average waiting time by increasing willingness to share from 0\% to 100\% while having 1000 less active drivers (scenario with $\omega=300$ seconds).

\begin{figure}[ht]
    \centering
    \includegraphics{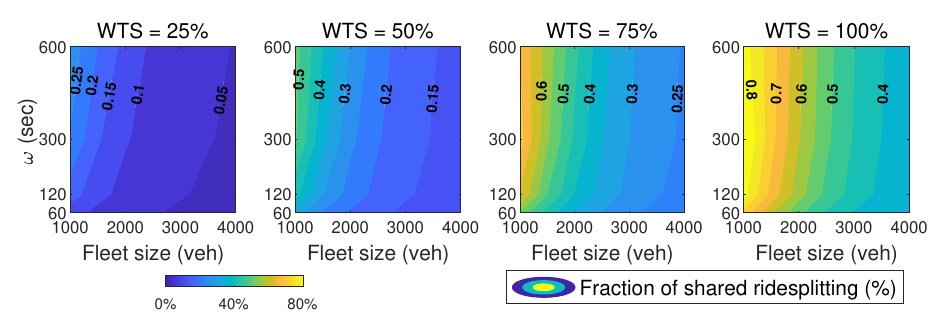}
    \caption{Fraction of shared rides from all rides as a function of fleet size, waiting time tolerance ($\omega$) and willingness to share (WTS).}
    \label{fig:multi_reg_shared}
\end{figure}

It is interesting how the sensitivity analysis shows the occurrence of shared rides.
Figure \ref{fig:multi_reg_shared} confirms that passengers' willingness to share is the most relevant parameter when computing the number of shared rides out of all provided ride-sourcing rides.
Waiting time tolerances become relevant only when too small, severely reducing the number of shared rides.
As one could expect from the assumption of matching incoming requests to the nearest driver and not prioritizing shared rides, larger fleets of active drivers decrease the number of shared rides.

\begin{figure}[ht]
    \centering
    \includegraphics[width=\textwidth]{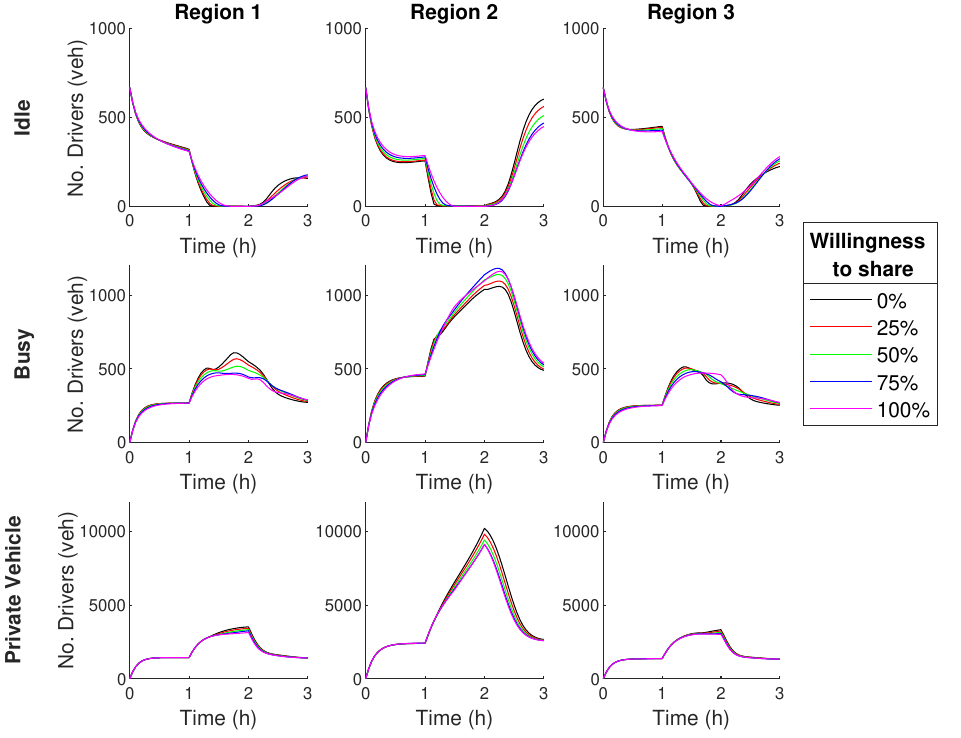}
    \caption{Summarized number of idle and busy drivers and private vehicles for the cases with 2000 ride-sourcing drivers and 10 minutes of waiting time tolerance.}
    \label{fig:multi_reg_dynamic}
\end{figure}

A close look at the dynamics of specific instances reveals how some parameters change drivers' activities in the experiment.
Figure \ref{fig:multi_reg_dynamic} shows that regions 1 and 2 remain nearly one hour without idle drivers, losing incoming ride-hailing requests.
Higher passengers' willingness to share was unable to avoid such a situation but still managed to serve more passengers, reducing the number of ride-sourcing travelers switching to private vehicles in all regions.
At the peak, the difference was around 1000 fewer private vehicles only in region 2, comparing scenarios of 0\% and 100\% willingness to share.

\section{Comparison with a detailed event-based simulator} \label{sec:num_results}

\begin{figure}[ht]
    \centering
    \includegraphics{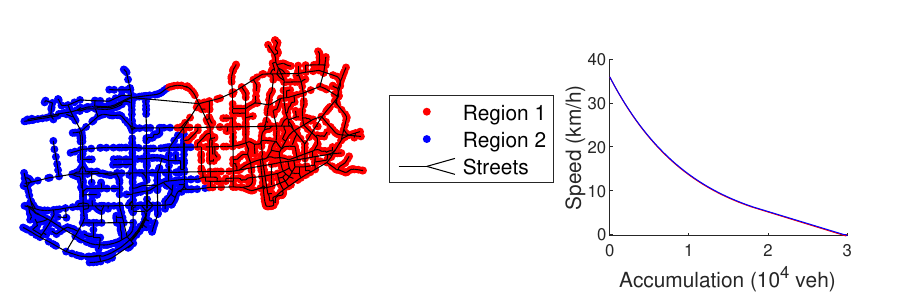}
    \caption{Shenzhen central business district separated in two regions and their respective Speed-MFDs used in the accuracy analysis.}
    \label{fig:2reg_data}
\end{figure}

The road network for the Futian and Luohu districts of Shenzhen, China, forms the background for the study.
The considered network consists of 1'858 intersections connected by 2'013 road segments.
In total, the Origin-Destination demand data contained around 200'000 requests collected from taxi operations using GPS coordinates \citep{ji_etal_2014}.
The experiment used a simulator based on \citet{beojone_geroliminis_2021}, which had historical data translated into Table \ref{tab:2reg_data} to use in the evaluated forecasts.
It also includes the MFD data in Figure \ref{fig:2reg_data} and a constant coefficient of variation $\sigma/L=0.57$.

\begin{table}[ht]
    \centering
    \caption{Two-region model parameters.}
    \begin{tabular}{ccccccccc}
        \hline
        \multirow{2}{*}{OD-pair} & \multirow{2}{15mm}{\centering{Demand ratio}} & \multicolumn{2}{c}{Transfer ratios ($\theta_{okd}$)} & \multicolumn{2}{c}{Passage ratio ($\beta^k_{od}$)} & \multicolumn{3}{c}{$L^\text{drop}_{K_{od}}$ (km)} \\
         & & $k=1$ & $k=2$ & $k=1$ & $k=2$ & $RH/PV$ & $S1$ & $S2$ \\
        \hline
        1-1 & 0.390 & n/d & n/d & 1 & 0 & 2.801 & 2.773 & 2.746 \\
        1-2 & 0.116 & 0   & 1   & 1 & 1 & 3.324 & 3.247 & 3.069 \\
        2-1 & 0.111 & 1   & 0   & 1 & 1 & 3.854 & 3.733 & 3.743 \\
        2-2 & 0.383 & n/d & n/d & 0 & 1 & 3.315 & 3.280 & 3.257 \\
        \hline
    \end{tabular}
    \label{tab:2reg_data}
\end{table}

\subsection{Simulation/Plant description}

The simulation/plant consists of an event-based spatial traffic simulation based in \citet{beojone_geroliminis_2021}.
It tracks every new trip based on its geographical origin, destination, and traveled distance in an urban network designed as a graph of roads and intersections.
Differently from classical trip-based models, vehicles have their microscopic geographical positioning tracked to evaluate detailed passenger-driver matching constraints for ride-hailing and ridesplitting.
A Speed-MFD estimates time-varying speeds shared among all links of a region.
The previous eliminates the expensive traffic assignment process, and vehicles may travel through the shortest path.
To have accurate positions and passenger-driver matching evaluations, each entity in the simulation has a tuple of information characterizing them.

The arrival of a passenger marks the start of a ride-sourcing request.
Waiting time and detour tolerances are set for all passengers.
The matching process for ride-hailing requires an idle driver close enough to the arriving passenger to comply with the waiting time tolerance.
For ridesplitting, besides idle drivers, those assigned to another ridesplitting passenger are potential assignments for arriving passengers.
However, in these cases, the evaluation must also check whether the detour will be acceptable for both passengers.
If a ride request is feasible, it is accepted and a ride-sourcing driver is assigned to pick-up and deliver the respective passenger.

Ride-sourcing drivers are responsible for picking up and delivering passengers in the modeled road network.
Tracking their positions and activities allows the simulation to check their availability for matching with arriving passengers and dispatching them accordingly.
The simulation follows TNCs' common practice of assigning passengers to the closest available driver on an FCFS basis.
Assignments determine the sequence of visited intersections for pick-ups and drop-offs.
In the case of ridesplitting requests, they are ordered to minimize the total traveled distance as long as the detour tolerance is fulfilled for all involved passengers.

The majority of entities affecting traffic is the background traffic.
A simpler tuple represents private drivers' situation and position.
Once the driver reaches the destination, the vehicle leaves the system (by parking outside the road space, for instance).
These entities do not interact with the ride-sourcing service, except for the lost ride-sourcing requests using private vehicles to fulfill their trip demand and traveling speeds.

Differently from \citet{beojone_geroliminis_2021}, we separated the studied area into a set of regions.
Moreover, both ride-sourcing drivers and private vehicles have additional properties to track their interregional path (including each intraregional trip length and sequence of regions in a trip).
The shortest paths (distance and sequence of intersections) are defined using a Floyd-Warshall algorithm.

\subsection{Error evaluation} \label{sec:error_comp}

In the error evaluation component, the simulation provides reference values, and the dynamic model provides the forecasts for comparison.
However, depending on the application, the model must provide predictions for different time horizons.
An MPC controller used for real-time fleet management needs several short-term predictions, and its efficiency relies on the quality of those \citep{sirmatel_geroliminis_2018a,sirmatel_geroliminis_2021}.

Therefore, we mimic an MPC controller using a rolling time horizon framework to evaluate the model.
Every $\Delta t$ time units, the simulation halts and describes the system, including information about ride-sourcing and private vehicle numbers and their respective remaining distances.
Then, from halt time $t_i$, the model forecasts the system's evolution for the next $T$ steps of $\delta t$ time units.
Note that halting times $t_i$ are $\Delta t$ units apart from each other and $T \delta t \geq \Delta t$.

Firstly, the evaluation computes the error related to estimates of state $K_{od}$ for $T$ forecast steps starting at $t_i$, called $\varepsilon^K_{od}(t_i,T \delta t)$.
Equation [\ref{Eq:error_ind}] illustrates the absolute error of the prediction for state $K_{od}$, measured as the ``number of vehicles'' (veh).

\begin{align}
    \varepsilon^K_{od}(t_i,T \delta t) = \sum^T_{l = 1} \left| \hat{n}^K_{od}(t_i, l \delta t) - n^K_{od}(t_i + l \delta t) \right| \label{Eq:error_ind}
\end{align}

\noindent
where $n^K_{od}(t)$ stands for the actual number of vehicles in state $K_{od}$; and $\hat{n}^K_{od}(t_i,t_i+\delta t)$ is the predicted value of $n^K_{od}(t_i+\delta t)$ when starting the prediction at $t_i$.

The relative error $\varepsilon(t_i, T \delta t)$ aggregates all errors for a given halting time $t_i$ and prediction horizon $T \delta t$.
The estimated error of the model is called $\varepsilon(T \delta t)$.
It summarizes the errors for all halting times $t_i$ depending on $T \delta t$.
Equations [\ref{Eq:subtot_error}] and [\ref{Eq:tot_error}] depict both of these dimensionless errors.

\begin{align}
    \varepsilon(t_i, T \delta t) = & \frac{\sum_{K \in \mathbb{S}}\sum_{o,d}\varepsilon^K_{od}(t_i, T \delta t)}{\sum_{K \in \mathbb{S}}\sum_{o,d}\sum_{l = 1}^T n^K_{od} (t_i + l \delta t)}, & \mathbb{S}=\{I, RH, S1, S2, PV\}, \mbox{ and } \{o,d\} \in \mathcal{R}^2 \label{Eq:subtot_error} \\
    \varepsilon(T \delta t) = & \sum_{t_i}\varepsilon(t_i, T \delta t), & t_i = 0,\Delta t, 2\Delta t, ..., t_f, \text{ and } T=1, 2, 3, 4, 5 \label{Eq:tot_error}
\end{align}

In this experiment, the simulation halts every $\Delta t = 3$ minutes ($0.05$h).
Then the model predicts from 1 to 5 steps ($T$) of 6 minutes ahead of time, completing up to 30 minutes of forecasts.
The settings for this experiment are illustrated by $\Delta t=0.05$h, $t_i=0, \Delta t, 2\Delta t, ..., 3$h, $\delta t=0.1$h, and $T=1, 2, ... 5$.
We refer to this experiment as the ``short forecast.''

Other applications (e.g. pricing) might require longer predictions.
Thus, it is necessary to understand its limitations and ability to describe system dynamics for different time horizons.

A second experiment consisted of a single model run for the entire evaluation period, receiving information from the simulation only at the beginning.
For the remaining time, there is no information exchange between the model and the plant.
In summary, the settings for this experiment has a single $t_i=0$, $\Delta t = 3$h, and a $T \delta t = 3$h.
We refer to this experiment as the ``long forecast.''

\subsection{Benchmark models}

To better emphasize the importance of having a more complex model for ride-sourcing dynamics compared to an accumulation-based MFD model, we utilized a benchmark model developed to model cruising for parking with MFD dynamics \citep{geroliminis_2015}.
It was one of the first efforts to integrate dynamic trip lengths and a state representation that decomposes the trip of a vehicle to various components, as required given the features of ride-sourcing trips, but in an accumulation-based model.
Simply speaking, the total production of the vehicles splits among the different states in a way analogous to the accumulations, and trip endings are estimated by dividing the specific production by the average trip length of the state (that can be time-dependent).
Moreover, the literature contemplates dynamic models for taxi and ride-hailing services, which form a relevant benchmark for the proposed model.
However, the presented model is distinctive for including ridesplitting operations deliberately.
Therefore, we aggregated all ride-sourcing activities into a single busy state for the benchmark model.
\citet{ramezani_nourinejad_2018} also had a similar activity description for private and taxi vehicles separated into dispatched and occupied ones.
In this approach, except for idle ride-sourcing vehicles, trip lengths are assumed constant and were estimated using the plant data.
Trip lengths considered the entire distance a ride-sourcing vehicle traveled from its assignment to a passenger until the last passenger drop (becoming idle again), independently of the service option.
We refer to this benchmark model as the ``Acc.-based model'' in the figures.

One could acknowledge that ride-sourcing operators are indifferent to traffic conditions when evaluating their service dynamics.
For this reason, we wanted to evaluate the impact of tracking traffic conditions during predictions.
To this end, we adapted this benchmark model to assume a constant free-flow speed $v_o(n_o(t)=0)$ at the prediction horizon.
In this case, we refer to it as the ``Benchmark No-traffic''.
The benchmark's purpose is to highlight that even if TNCs might not be interested in the congestion their operations create, they should account for it if they are interested in managing their quality of service with real-time strategies (e.g., repositioning or surge pricing).

\subsection{Model evaluation}

The initial evaluation of the proposed model (prediction quality and stability, without measuring errors) consists of the experiments called ``long forecast'' and ``short forecast,'' described in Section \ref{sec:error_comp}.

\begin{figure}[ht]
    \centering
    \includegraphics[width=\textwidth]{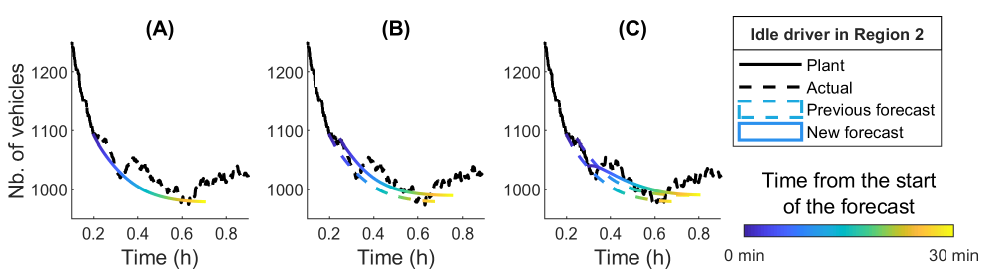}
    \caption{Rolling time horizon prediction instances for idle ride-sourcing vehicles in Region 2. Forecasts starting 12 (A), 15 (B) and 18 (C) minutes from the experiment beginning.}
    \label{fig:forecast_zoom}
\end{figure}

To illustrate the experiment mimicking an MPC controller (``short forecast''), Figure \ref{fig:forecast_zoom} shows three consecutive steps computed in a rolling time horizon prediction for idle ride-sourcing drivers in Region 2.
The model predicts future system conditions every 3 minutes (0.05 hours).
The model only considers the first 10 time steps (completing 30 minutes of forecasts) to reasonably use computational resources.
The feedback loop from the plant to the prediction model estimates system states, including ride-sourcing and private vehicle information.

\begin{figure}[ht]
    \centering
    \includegraphics[width=\textwidth]{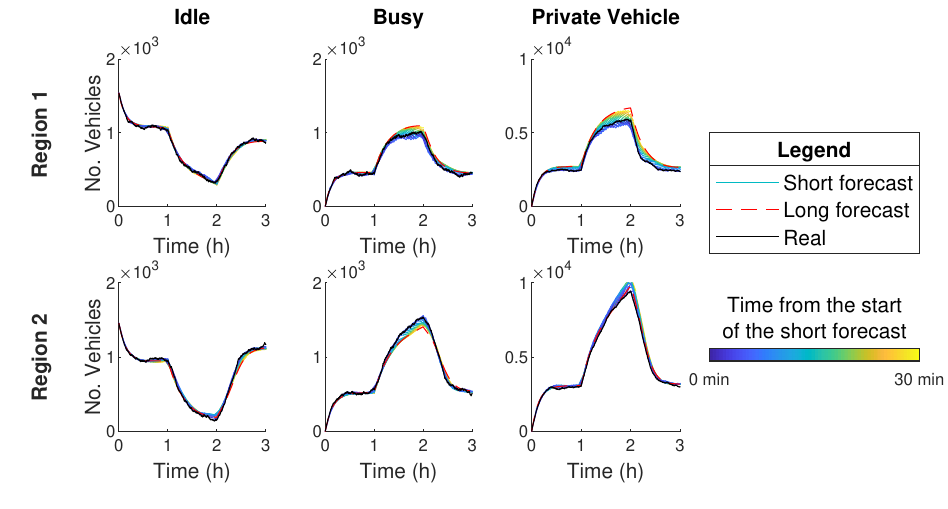}
    \caption{Results for long and short forecasts compared directly to the plant results. Model states are aggregated per current region and vehicle situation.}
    \label{fig:forecasts}
\end{figure}

To assess whether the model can capture the state dynamic evolution, we separated a single round of the plant with the respective forecasts for the number of vehicles.
Figure \ref{fig:forecasts} depicts these measurements for short and long forecasts.
Note that, for visualization purposes, we aggregated the number of vehicles according to their current region.
We also aggregated all ride-sourcing vehicles with at least one assigned passenger into a ``Busy'' classification.
In general, estimated values followed plant values closely for most states.
In some short forecasts, such as $S1_{12}$, $PV_{11}$, $PV_{21}$, and most $S2_{od}$, the model initially moves away from the plant data, but it returns to the values close to the ``long forecasts'' and the plant values.
The deviations remained, at most, in the order of $10^1$ for ride-sourcing and $10^2$ for private vehicles.
The previous highlights that examining short forecasts may provide a better test of robustness.
The regional number of idle and busy drivers is crucial for ride-sourcing operations. Various strategies require those. Some examples are vacant vehicle relocation, surge pricing, integration of ride-sourcing in High-Occupancy-Vehicle or High-Occupancy-Toll lanes, and perimeter control.

\begin{figure}[ht]
    \centering
    \includegraphics{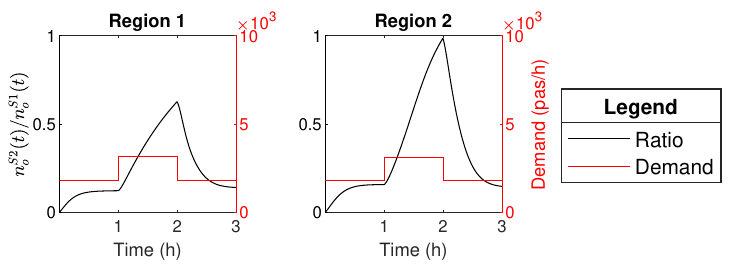}
    \caption{Ratio of shared rides compared to ridesplitting demand.}
    \label{fig:forecast_split}
\end{figure}

One of the central concerns in modeling ridesplitting activities is ensuring the model can capture key service characteristics, which are state- and demand-dependent.
Figure \ref{fig:forecast_split} depicts the ratio of drivers $n^{S2}_{od}(t) / n^{S1}_{od}(t)$ on ridesplitting activities in each region.
As one would expect, the more passengers joining ridesplitting, the more passengers have shared rides.
In the beginning, few drivers carry multiple riders since, most times, an idle vehicle is the closest one to arriving passengers.
However, once the demand grows, it forms the pool of drivers with a single ridesplitting passenger allowing for more shared rides, where hundreds of drivers in both regions have two simultaneous passengers.
For instance, in Region 2, the number of drivers carrying two passengers almost equals the number of drivers carrying a single passenger at 2h.
It shows a seven-fold increase in a period of 75\% larger demand, highlighting the responsiveness to market thickness.

\subsection{Benchmark comparison}

\begin{figure}[ht]
    \centering
    \includegraphics{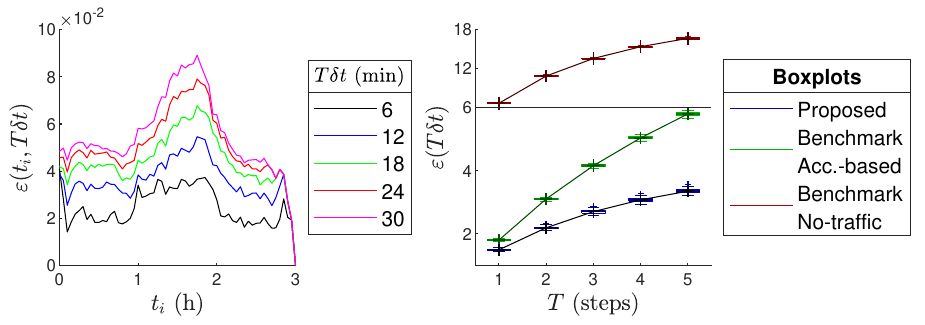}
    \caption{Summary of error measurements. (Left) Subtotal forecast errors for different $T \delta t$; and (Right) Boxplots of total errors according to the $T \delta t$.}
    \label{fig:errors}
\end{figure}

Error measurements can provide a detailed analysis of the quality of the forecasts and how they deteriorate at later steps.
In Figure \ref{fig:errors}, we compile the forecasts from 30 independent experiment runs.
Firstly, in Figure \ref{fig:errors} (Left), one can note that relative errors are naturally higher for longer predictions, as expected.
Furthermore, errors often remained below 5\%, even on forecasts of 12 minutes or more.
Errors were higher than 6\% only at the most crowded moments.
Measurements on $\varepsilon(t_i,T \delta t)$ converge at 0 by the end of the experiment because there are no forecasts (nor plant data) after 3h.
In Figure \ref{fig:errors} (Right), we included total error measurements for all the benchmarks and the proposed model.
Excluding `Benchmark No-traffic', boxplots presented increasing variability in later forecast steps but remained small compared to the average.
For instance, coefficients of variation ranged between 0.005 and 0.022 for the benchmark and the proposed models, respectively.
Total error measurements, $\varepsilon(T \delta t)$, presented nearly linear increases for the number of steps.
Errors of the `Acc.-based MFD model' were nearly double those of the proposed model, while the `Benchmark No-traffic' model marked around $5$ times higher errors (plotted above the other models).

We acknowledge that we computed error measurements in Figure \ref{fig:errors} for all model states, not only the aggregated ones in Figure \ref{fig:forecasts}.
Hence, errors for the proposed model accounted for all 18 states, while the `Acc.-based MFD model' and the `Benchmark No-traffic' accounted for only 10 states.
Total and subtotal errors of the proposed model were inferior to those of all benchmarks.
It confirms the proposed model as a better approximation to the traffic system.

\begin{figure}[ht]
    \centering
    \includegraphics{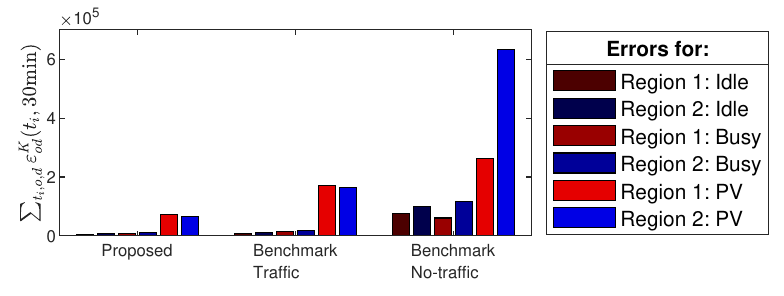}
    \caption{Summary of total errors in forecast of 30 minutes (5 steps). Total errors per vehicle situation and current region.}
    \label{fig:error_pie}
\end{figure}

To evaluate if the errors are distributed differently in each model, we separated them for all vehicles according to their situation and current region in Figure \ref{fig:error_pie}.
To have a fair comparison, we aggregated the vehicles in groups before computing the errors ($\varepsilon^K_{od}(t_i,T \delta t)$) in all models.
As shown in Figure \ref{fig:error_pie}, private vehicles represent most of the errors for all models.
In the `Acc.-based MFD model,' the errors have a similar share distribution as the demand, where 85\% of it refers to private vehicles.
The proposed model reduced the errors for these vehicles by 60\%, even with the same modeled states.
On the other hand, the proposed model has 8 states more for ride-sourcing vehicles, and errors reduce by between 28\% and 42\%.
The total error dropped to less than half, indicating that the computation of trip completion and transfer flows is responsible for such results.
At the same time, it provided more detailed information on ridesplitting operations.
As the last evaluation on the dynamics of `Benchmark No-traffic,' we checked whether the increased errors from Figure \ref{fig:errors} concentrated in private vehicles.
However, errors had a similar distribution as the other tested models, highlighting the importance of traffic dynamics in evaluating ride-sourcing operations (even if the operator is not interested in it).

As a note, we ran tests separating pick-up and delivery activities to evaluate possible shortcomings of the aggregation process.
However, even with the additional description, it had slightly higher $\varepsilon(T \delta t)$ and $\varepsilon^K_{od}(t_i,30\text{min})$ than the proposed model.
Results were worse than the proposed model because it is more susceptible to measurement noises, particularly those in pick-up activities. After all, they have very short average trip lengths.

\section{Conclusions} \label{sec:conclusions}

We proposed a dynamic model capable of representing ride-sourcing services and private vehicles macroscopically in an urban network separated in a multi-region setting.
We supplied the processes for estimating the required parameters and computation of errors.
It depicted mass conservation equations for both ride-sourcing and private vehicles (backgrgound traffic).

We evaluated a multi-region setting and its sensitivity to passengers' willingness to share, their waiting time tolerance, and ride-sourcing drivers' fleet size.
The model directly captured the effects of wild-goose-chase \citep{castillo_etal_2018}, confirming that limiting matching radius -- in this case, represented by decreasing waiting time tolerance -- hinders its effects, decreasing the overall waiting time and number of lost requests, even in transient scenarios \citep{xu_etal_2020,yang_qin_ke_ye_2020}.
The multi-regional setting could further limit pick-up distances by limiting the matching radii, which could be upper bounded by well-known methods based on the region area and street network topology (see \citet[Chapter 3]{larson_odoni_1981}, for instance).

Assuming an FCFS matching scheme, the model showed the fleet size and willingness to share as crucial to match travelers into a single vehicle.
Forming a pool of passengers and having fewer drivers to serve them forces the operator to bring travelers together.
\citet{beojone_geroliminis_2021} reported similar findings, which evaluated a detailed simulation instead of a dynamic model.
Furthermore, observing the regional dynamics further allowed identifying the influence of demand patterns on local driver availability and instantaneous traffic conditions represented primarily by the background traffic of private vehicles.

The final evaluation tested the model quality to five different prediction horizons, which presented increasing errors to their lengths but remained below 10\% in all cases.
In the next step, we compared the proposed model to different benchmarks, and the computed errors were only a fraction of theirs.
Like the exceptional error measurements, the actual values closely followed the plant at all times in a test built to mimic the rolling time-horizon structure of an MPC controller.
Moreover, the proposed model had lower error measurements than the benchmarks in all states.
Representing a unique state encompassing pick-up and delivery activities for each ride-sourcing service decreased the sensitivity of the proposed model to measurement noises.
Precise predictions for ride-sourcing in dynamically congested areas present the next step towards better traffic control and service operations management.

We performed additional tests on the model, evaluating its sensitivity to noises in the inputs of private vehicles.
However, for scenarios with an average noise of 15\% or larger, total errors decreased at longer prediction horizons.
Noises were limited to measurements of the starting number of private vehicles, assuming that, with current technologies, one can have all information needed from ride-sourcing vehicles.
Even in a scenario with advanced technology employed in collecting precise traffic data, stability and robustness must be regarded as vital characteristics of models when constructing real-life solutions.
Based on our findings, besides the mobility benefits of ridesplitting, it also increases model stability, decreasing the sensitivity to noise.
In light of such findings, as models' complexity can increase, careful and systematic analysis of prediction errors and sensitivity to noises in particular measurements pose an unavoidable step for model evaluations.

This paper is among the first attempts to present and evaluate such a model with ridesplitting (shared rides) in the literature.
The success of a dynamic model with ride-sourcing provides scholars, practitioners, and authorities a tool for measuring the interactions such services over traffic in a simulated and fast environment and proper for studies improving shared rides in a congested urban area.
Further developments on repositioning strategies ride-sourcing services with ridesplitting options and traffic congestion can profit from dynamic models such as the one proposed in this paper.
Notably, the proposed model could support high-level repositioning decisions in hierarchical problems like the one seen in \citet{yildirimoglu_etal_2018}.
Furthermore, problems in strategic market/regulatory responses to ride-sourcing services in transient situations can employ the proposed model.

\appendix

\section{List of notation}

\begin{table}[ht]
    \centering
    \caption{List of notation with brief description.}
    \label{tab:notation}
    \begin{tabular}{cp{14cm}}
        \hline
        Notation & Description \\
        \hline
        $\mathcal{R}$   & Set of modeled regions. \\ [2pt]
        $\mathcal{R}_{o}$   & Adjacent regions to region $o$. \\ [2pt]
        $\theta_{okd}$ & Ratio of trips from $o$ to $d$ where region $k$ is the next on the path. \\ [2pt]
        $n^K_{od}(t)$ & Number of vehicles in a state $K$, currently at region $o$, with destination in region $d$. \\ [2pt]
        $M^K_{od}(t)$ & Total remaining distance that vehicles in state $K$, currently at region $o$, with a destination in region $d$ must travel. \\ [2pt]
        $v_o(t)$ & Instantaneous average speed at region $o$ \\ [2pt]
        $O^K_{od}(t)$ & Instantaneous trip completion rate of vehicles in a state $K$, currently at region $o$, with a destination in region $d$.\\ [2pt]
        $\hat{O}^{S1}_{od}(t)$ & Estimated trip completion rate of single ridesplitting drivers without interruption.\\ [2pt]
        $L^K_{od}(t)$ & Average trip length for vehicles entering state $K$ with an origin region $o$ and destination region $d$.\\ [2pt]
        $L^*_{od}(t)$ & Average remaining distance to be travelled in steady-state for drivers state $K$, currently in region $o$ and destination in region $d$.\\ [2pt]
        $\sigma^K_{od}(t)$ & Standard deviation of $L^K_{od}(t)$.\\ [2pt]
        $\alpha$ & M-model parameter related to the trip length distribution (set to $-3$).\\ [2pt]
        $L^\text{pick}_{K_{od}}(t)$ & Instantaneous average pick-up trip length for vehicles in state $K$ for trips with origin in region $o$ and destination in region $d$.\\ [2pt]
        $L^\text{drop}_{K_{od}}(t)$ & Instantaneous average delivery trip length for vehicles in state $K$ for trips with origin in region $o$ and destination in region $d$.\\ [2pt]
        $R$ & Acceptable pick-up distance radius.\\ [2pt]
        $\omega$ & Passengers' waiting time tolerance.\\ [2pt]
        $pl^s_{od}(t)$ & Loss probability of travellers for service $s$ with origin region $o$ and destination region $d$.\\ [2pt]
        $\lambda^s_{od}(t)$ & Travellers' arrival rate for service $s$ with origin region $o$ and destination region $d$.\\ [2pt]
        $\bar{\lambda}^s_{od}(t)$ & Travellers' entering rate for service $s$ with origin region $o$ and destination region $d$.\\ [2pt]
        $\rho^s_{od}(t)$ & Fraction of entering rides of service $s$ with origin region $o$ and destination region $d$, assigned to idle drivers. \\ [2pt]
        $n^{av,\text{LIFO}}_{od}(t)$ & Number of available drivers for a ridesplitting ride from $o$ to $d$, served in a LIFO trip scheme. \\ [2pt]
        $n^{av,\text{FIFO}}_{od}(t)$ & Number of available drivers for a ridesplitting ride from $o$ to $d$, served in a FIFO trip scheme. \\ [2pt]
        $\beta_{od}^h$ & Ratio of trips in $o$ that will pass through region $h$ anytime on the path to $d$. \\ [2pt]
        $\vartheta_{ood}(t)$ & Fraction of $S2_{od}$ rides delivering one of the passengers in the current region $o$. \\ [2pt]
        $\varepsilon^K_{od}(t_i,T \delta t)$ & State-wise error of a prediction started at $t_i$, for $T$ steps of $\delta t$ duration. \\ [2pt]
        $\varepsilon(t_i, T \delta t)$ & Subtotal error of a prediction started at $t_i$, for $T$ steps of $\delta t$ duration.\\ [2pt]
        $\varepsilon(T\delta t)$ & Total error of predictions with $T$ steps of $\delta t$ duration. \\ [2pt]
        \hline
    \end{tabular}
\end{table}

\section{Monte Carlo Simulation} \label{sec:monte_carlo}

A Monte Carlo simulation evaluated the influence on the non-acceptance (loss) of incoming requests after checking the feasibility constraints (waiting time and/or detour) of factors such as the available fleet size $n^{av}_{od}(t)$, the regional average speed $v_o(t)$, the waiting time tolerance $\omega$, and the Idle-Busy drivers' ratio $\rho^s=n^I(t)/n^{av}_{od}(t)$ (Equation [\ref{Eq:gamma}]).
We simulated loss probabilities for the variable tuple $(n^{av}_{od}(t),v_o(t),\omega,\rho^s)$.
Where we tested all possible combinations of the following values for each variable:

\begin{itemize}
    \item $v_o \in \{5, 10, 15, 20, 25, 30, 35, 40\}$ (unit: km/h);
    \item $n^{av}_{od} \in \{10, 30, 60, 110, 150, 210, 270\}$ (unit: $\#$veh);
    \item $\omega \in \{2, 5, 8, 11, 14, 17, 20\}$ (unit: min); and
    \item $\rho^s \in \{0.1, 0.3, 0.5, 0.7, 0.9, 1\}$ (unitless).
\end{itemize}

\noindent
Note that $\rho^s=1$ will make the problem indifferent in terms of service since all drivers are idle.

The simulation considers a street network divided into regions, depending on the desired case.
Since drivers and passengers must be in the same region to perform an assignment, each region had an independent Monte Carlo simulation.
Under the assumption of an FCFS discipline without batching, it always yields a higher supply of drivers than demand for passengers.\footnote{%
See Footnote \ref{foot:batch}.}
Moreover, separate simulations evaluated ride-hailing and ridesplitting services.
Algorithm \ref{alg:monte_carlo} shows a pseudo-code that illustrates the Monte Carlo simulation.
We sample and test numerous potential passengers ($pasSamp=500$) individually, considering all vehicles in the sampled fleet.
Furthermore, to ensure that there is no bias in the vehicle positioning, we sample several combinations of positions for drivers ($vehSamp=20$) for each set of parameters.
Four properties describe each of the sampled drivers:

\begin{itemize}
    \item Current position ($.curr$): Node in the street network;
    \item Origin position ($.orig$): Node in the street network -- empty for idle drivers;
    \item Destination position ($.dest$): Node in the street network -- empty for idle drivers;
    \item Busy flag (boolean): `false' for idle drivers, `true' for busy ones;
\end{itemize}

\begin{algorithm}
    \linespread{1.15}\selectfont
    \SetKwFor{ForEach}{for each}{do}{end}
    \KwData{$allF$, \hspace{53pt}\texttt{// Set of all tested fleet sizes} \newline
            $allR$, \hspace{53pt}\texttt{// Set of all tested Idle-Busy ratios} \newline
            $allV$, \hspace{53pt}\texttt{// Set of all tested traveling speeds} \newline
            $allW$, \hspace{50pt}\texttt{// Set of all tested waiting time tolerances} \newline
            $vehSamp=20$, \hspace{5pt}\texttt{// Number of sampled sets of positions for drivers.} \newline
            $pasSamp=500$, \texttt{// Number of sampled passengers per test.}}
    \KwResult{$pl^s_{od}$ \hspace{21pt}\texttt{// Loss probabilities.}}
    $auxPL \gets \text{5-D array of zeros}$\;
    \ForEach{$idxF \in \{1,...,|allF|\}$}{
        $f \gets allF(idxF)$\;
        \ForEach{$idxR \in \{1,...,|allR|\}$}{
            $r \gets allR(idxR)$\;
            $f_{Idle} \gets f * (1-r)$\;
            $f_{Busy} \gets f * r$\;
            \ForEach{$idxP \in \{1,...,\text{\normalfont{vehSamp}}\}$}{
                Initialize empty object $VEH$ with properties $.orig$, $.dest$, $.curr$, $.busy$\;
                Sample $f_{Idle}$ Idle drivers into object $VEH$\;
                Append a sample of $f_{Busy}$ Busy drivers into object $VEH$\;
                Sample a set of $pasSamp$ OD-pair into an object $PAS$\;
                \ForEach{$idxV \in \{1,...,|allV|\}$}{
                    $v_o \gets allV(idxV)$\;
                    \ForEach{$idxW \in \{1,...,|allW|\}$}{
                        $\omega \gets allW(idxW)$\;
                        match $\gets matdis(VEH,PAS,v_o,\omega)$\tcp*[r]{Check match constraints}
                        \tcc{Match is a vector with $pasSamp$ logical elements \newline
                        -- $0$ indicates that a passenger is NOT LOST. \newline
                        -- $1$ indicates that a passenger is LOST.}
                        $auxPL(idxF,idxR,idxP,idxV,idxW) \gets mean(\text{match})$\;
                    }
                }
            }
        }
    }
    $pl^s_{od} \gets $ Average of $auxPL$ over the 3rd dimension \tcp*[r]{over $vehSamp$ tests.}
    \Return{$pl^s_{od}$}  \vspace{4pt}\;
\caption{Monte Carlo algorithm pseudo-code.} \label{alg:monte_carlo}
\end{algorithm}

The evaluation of the waiting time and detour tolerances are described below in Algorithm \ref{alg:matdis} to detail the matching constraints evaluated in the function in Algorithm \ref{alg:monte_carlo}.
In this process, besides the main parameters, we use the shortest distance between two points in the street network $allDist(\cdot,\cdot)$, previously computed using a Floyd-Warshall algorithm, and a maximum detour tolerance $detour$.
Note that the evaluated constraints for ride-hailing and ridesplitting are the same as in \citep{beojone_geroliminis_2021}.

\begin{algorithm}
    \linespread{1.15}\selectfont
    \SetKwFor{ForEach}{for each}{do}{end}
    \SetKwFunction{FMain}{$matdis$}
    \SetKwProg{Fn}{function}{:}{}
    \KwData{$allDist$, \texttt{// Matrix with the shortest distance for every OD-pair.} \newline
            $detour$, \hspace{4pt}\texttt{// Detour tolerance.}}
    \Fn{\FMain{$VEH$, $PAS$, $v_o$, $\omega$}}{
        match $\gets$ vector with $|PAS|$ elements all equal to `true'\;
        \ForEach{pas $\in PAS$}{
            $k \gets $ true\;
            $veh \gets 0$\;
            \While{$k$ {\normalfont{is true} \textbf{\&\&}} \textit{veh}$\leq |VEH|$}{
                $veh \gets veh + 1$\;
                \If{$allDist(VEH(veh).curr,PAS(pas).origin)/v \leq \omega$}{
                    \eIf{$VEH(veh).busy==true$}{
                        match(pas) $\gets$ false\;
                        $k \gets $ false\;
                    }{
                        $dist01 \gets allDist(VEH(veh).orig,VEH(veh).dest) * (1+detour)$\;
                        $dist02 \gets allDist(pas.orig,pas.dest) * (1+detour)$\;
                        $dist11 \gets allDist(VEH(veh).orig,VEH(veh).curr) + allDist(VEH(veh).curr,pas.orig) + allDist(pas.orig,VEH(veh).dest)$\;
                        $dist12 \gets allDist(pas.orig,VEH(veh).dest) + allDist(VEH(veh).dest,pas.dest)$\;
                        $dist21 \gets allDist(VEH(veh).orig,VEH(veh).curr) + allDist(VEH(veh).curr,pas.orig) + allDist(pas.orig,pas.dest) + allDist(pas.dest,VEH(veh).dest)$\;
                        $dist22 \gets allDist(pas.orig,pas.dest)$\;
                        \tcp{Test ridesplitting (two possible sequences).}
                        \If{$dist11 \leq dist01$ {\normalfont\textbf{\&\&}} $dist12 \leq dist02$}{
                            match(passenger) $\gets$ false\;
                            $k \gets $ false\;
                        }
                        \If{$dist21 \leq dist01$ {\normalfont\textbf{\&\&}} $dist22 \leq dist02$}{
                            match(passenger) $\gets$ false\;
                            $k \gets $ false\;
                        }
                    }
                }
            }
                \If{vehicle \normalfont{pick-up time} $\leq \omega$ \&\& respecting possible detour constraints}{}
        }
        \KwRet match \vspace{4pt}
    }
\caption{Passenger-Driver matching test pseudo-code.} \label{alg:matdis}
\end{algorithm}

\section{Loss probability function estimation} \label{sec:linear_model}

As presented earlier, Equation [\ref{Eq:loss_prob_form}] is a function of the available fleet size $n^{av}_{od}(t)$, the regional average speed $v_o(t)$, the waiting time tolerance $\omega$, and the ratio of Idle-Busy drivers $\rho^s=n^I(t)/n^{av}_{od}(t)$.
Hence, we can linearize Equation [\ref{Eq:loss_prob_form}] into Equation [\ref{Eq:linearized_loss}], which allows us to estimate the coefficients to its equivalent linear regression model in Equation [\ref{Eq:linear_model}] (written using Wilkinson notation) using the least-square fit.

\begin{align}
    \log(-\log(pl^s_{od})) = & \ \log \gamma_0 + \gamma_1 \log n^{av}_{od} + \gamma_2 \log v_o + \gamma_3 \log \omega + \gamma_4 \log \rho^s \label{Eq:linearized_loss} \\
    Y \sim & \ I + n^{av}_{od} + v_o + \omega + \rho^s \label{Eq:linear_model}
\end{align}

\begin{figure}[ht]
    \centering
    \includegraphics{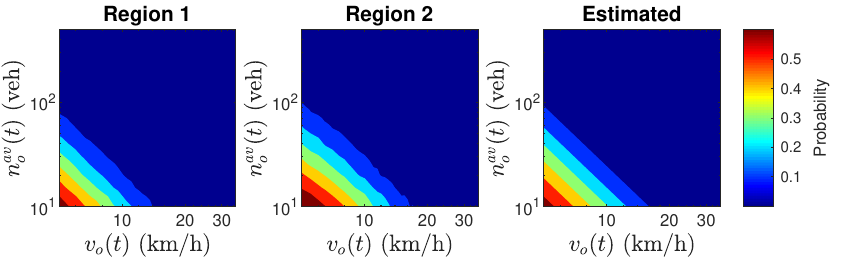}
    \caption{Estimated loss probability for passengers (both axes in log-scale): (a) Monte Carlo Simulation for region 1, (b) Monte Carlo Simulation for region 2, (c) Fitted equation results. Where $n^{av}_o(t)$ is the fleet of available vehicles at time $t$, while $v_o(t)$ is the instantaneous speed in the region $o$ at time $t$. The values for waiting time tolerance $\omega$ and the ratio of Idle-Busy drivers $rho^s$ were fixed at $10$ minutes and $1$, respectively.}
    \label{fig:loss_prob}
\end{figure}

For the sake of illustration, Figure \ref{fig:loss_prob} shows an instance of the fitted function in comparison with the simulated data of the two-region setting.

\bibliographystyle{cas-model2-names}

\bibliography{refs}

\end{document}